\documentclass[11pt,oneside,letterpaper]{article}
\usepackage{amssymb}
\usepackage{amsmath}
\usepackage[dvips]{graphicx}
\usepackage{setspace}
\usepackage{amsfonts}
\usepackage{fancyhdr}
\usepackage{xcolor}
\usepackage{graphicx}
\usepackage{rotating}
\usepackage{comment}
\usepackage{color}
\usepackage{cite}
\usepackage{braket}
\usepackage{subcaption}

\definecolor{darkgreen}{rgb}{0,0.5,0}
\definecolor{darkblue}{rgb}{0,0,0.6}
\definecolor{purple}{rgb}{0.4,.2,0.7}
\newcommand{\p}{\partial}

\newcommand{\f}{\frac}

\newcommand{\be}{\begin{equation}}
\newcommand{\ee}{\end{equation}}

\newcommand{\tp}{\tilde{\phi}_r}
\newcommand{\tc}{\tilde{\chi}}
\newcommand{\te}{\tilde{\eta}}

\usepackage[colorlinks=true,citecolor=darkgreen,linkcolor=black,urlcolor=purple]{hyperref}

\usepackage{pdfsync}

\makeatletter
\newcommand*{\defeq}{\mathrel{\rlap{%
                     \raisebox{0.3ex}{$\m@th\cdot$}}%
                     \raisebox{-0.3ex}{$\m@th\cdot$}}%
                     =} 
\makeatother

\def\be{\begin{eqnarray}}
\def\ee{\end{eqnarray}}

\newcommand{\bea}{\begin{eqnarray}}
\newcommand{\eea}{\end{eqnarray}}
\def\ben{\begin{equation}}
\def\een{\end{equation}}

\let\a=\alpha \let\b=\beta \let\g=\gamma \let\d=\delta 
   
 \let\m=\mu \let\n=\nu  \let\p=\phi 
\let\s=\sigma

  \let\D=\Delta  \let\L=\Lambda
\let\X=\Xi

\let\f=\frac

\def\be{\begin{equation}}
\def\ee{\end{equation}}
\def\ba{\begin{array}}
\def\ea{\end{array}}

\def\ba#1\ea{\begin{align}#1\end{align}}
\def\bs#1\es{\begin{split}#1\end{split}}

\renewcommand{\p}{\partial}

\allowdisplaybreaks  

\numberwithin{equation}{section}

\begin{document}
\onehalfspacing

\begin{center}

~
\vskip5mm

{\LARGE  {
Encoding beyond cosmological horizons \\ in de Sitter JT gravity
\\
\ \\
}}

Adam Levine$^1$ and Edgar Shaghoulian$^2$

\vskip5mm
{\it $^1$Institute for Advanced Study, Princeton, NJ 08540, USA\\
 $^2$ Department of Physics and Santa Cruz Institute for Particle Physics,\\
University of California Santa Cruz, Santa Cruz, CA 95064, USA
} 

\vskip5mm

{\tt arlevine@ias.edu, eshaghoulian@ucsc.edu}

\end{center}

\vspace{4mm}

\begin{abstract}
\noindent
Black hole event horizons and cosmological event horizons share many properties, making it natural to ask whether our recent advances in understanding black holes generalize to cosmology. To this end, we discuss a paradox that occurs if observers can access what lies beyond their cosmological horizon in the same way that they can access what lies beyond a black hole horizon. In particular, distinct observers with distinct horizons may encode the same portion of spacetime, violating the no-cloning theorem of quantum mechanics. This paradox is due precisely to the observer-dependence of the cosmological horizon -- the sharpest difference from a black hole horizon -- although we will argue that the gravity path integral avoids the paradox in controlled examples.
 \end{abstract}

\pagebreak
\pagestyle{plain}

\setcounter{tocdepth}{2}
{}
\vfill

\ \vspace{-2cm}
\renewcommand{\baselinestretch}{1}\small
\tableofcontents
\renewcommand{\baselinestretch}{1.15}\normalsize

\section{Introduction}
One of the central lessons of the past few years is that the semiclassical gravitational path integral knows about the encoding of the interior of the black hole in its Hawking radiation. Black hole horizons are ubiquitous in our universe, as they are believed to exist at the center of almost every galaxy. Perhaps even more ubiquitous is the cosmic horizon. Unlike a black hole, this horizon surrounds us, but similar to a black hole, it is believed to Hawking radiate at a characteristic temperature set by the size of the horizon. Furthermore, cosmic horizons have a thermodynamic entropy \cite{Gibbons:1976ue}
\be\label{gh}
S = \f{A}{4G}\,,
\ee
given by the same formula as the black hole entropy. Trying to understand the encoding of spacetime beyond the cosmic horizon from the finite cavity within it is a difficult problem; what would help is an ``exterior" or ``bird's eye" view of cosmology. Assuming an exit from inflation \cite{Freivogel:2006xu, Chen:2020tes, Hartman:2020khs} or fixing future boundary conditions at $\mathcal{I}^+$ \cite{Maldacena:2002vr, Strominger:2001pn} are closely related versions of providing this exterior view. They provide us with an infinite Hilbert space in which we can make arbitrarily precise measurements and therefore put the problem on a more similar footing to that of black holes.\footnote{The finite dimensionality of the Hilbert space for a de Sitter universe was first proposed in \cite{Bousso:1999dw, Banks:2000fe, Fischler} and developed in \cite{Bousso:2000nf, Banks:2001yp, Parikh:2002py, Dyson:2002nt,   Banks:2002wr,  Banks:2003pt, Banks:2006rx, Banks:2015iya, Banks:2018ypk, Banks:2020zcr, Susskind:2021yvs, Shaghoulian:2021cef}. For a study of fixed future boundary conditions see \cite{Anninos:2011jp}.} 

In this paper, we will take a similar exterior view of cosmology. Our model will be Jackiw-Teitelboim gravity with a positive cosmological constant, which has de Sitter space as a solution. It also has a black hole in de Sitter space as a solution, which can be thought of as a dimensional reduction of the Nariai black hole. Like in higher dimensions, this black hole solution admits an arbitrarily large analytic extension, which can have as many black holes and as many inflating regions as one desires. This analytic extension is often assumed to be a mathematical curiosity, but we will see that it passes some consistency checks. It will allow us to formulate our paradox, which we briefly outline below. 

In our spacetime with inflating regions separated by black hole regions, we will consider two observers -- Alice and Bob -- in distinct inflating regions. We will assume these observers' local patches have exited from inflation, and they are in a region where gravity is weak, such that the spacetime background can be fixed. We will refer to such regions where the spacetime geometry is fixed as \emph{frozen}.\footnote{Note that because gravity is turned off in these frozen regions, we can really think of Alice and Bob's Hilbert spaces as tensor factorizing. This should be contrasted with the idea that when the different patches are weakly gravitating there is a single, non-factorizing Hilbert space which represents them both.} This will be our exterior view. A picture is provided below in figure \ref{threeregions}. The theory governing where Alice and Bob live is some quantum field theory (QFT) on a curved background; the gravitational part of the spacetime simply prepares an initial state for the evolution of the QFT. Since the two QFTs are spacelike separated, all operators in Alice's system commute with all operators in Bob's system. But notice that the exteriors of Alice's and Bob's horizons overlap. If the exterior of Alice's horizon is encoded by the data in her cavity, and similarly with Bob, that means that both Alice and Bob encode the same piece of spacetime. This violates the no-cloning theorem, since they can both independently extract a bit from beyond the horizon without affecting each others' ability to do so. We will elaborate on this paradox in section \ref{sec:paradox}. 

Our resolution to this paradox, which we will describe in more depth in section \ref{sec:resolution}, will be that whether or not the geometry in Bob's distinct inflating region is taken to be frozen or not can have a drastic effect on Alice's ability to reconstruct operators in Bob's region (and vice versa for Bob). Furthermore, we will find that for most ``natural'' choices of state on the two asymptotic regions, the dominant saddle in the semi-classical path integral is one where Alice and Bob's regions exist in their own, disconnected spacetimes. In this case, each observer only encodes the region beyond their horizon \emph{within their connected portion of the universe}. Thus there is no overlap and no violation of no-cloning. In order to make the dominant saddle the one hosting both Alice and Bob in the same connected universe, one must first change the path integral prescription (act with an operator) which entangles the two asymptotic regions. In that case, both Alice and Bob will find the microscopic state of their inflating regions to be mixed and they will not be able to encode each other's regions.

What we are describing is similar to a ``time-like homology constraint'', which disallows consideration of entanglement wedges which are in the past of a portion of frozen spacetime in the \emph{semi-classical} saddle. The point of the present work will be to justify this constraint in an explicit example by illustrating how the Euclidean path integral disallows such quantum extremal surface saddles from contributing.

We now briefly outline the paper. In section \ref{sec:setup}, we will describe the set-up of JT gravity coupled to 2d conformal matter. We will also describe the analytically extended nearly-Nariai geometries with multiple black hole and cosmological horizons. We will describe how quantum corrections are important for understanding this spacetime.\footnote{Analytic extensions of these near-Nariai geometries were recently discussed in \cite{Aguilar-Gutierrez:2021bns}, but the authors in that work did not account for the backreaction due to quantum matter, which we include. We will also discuss the relevant boundary conditions and how to compute physical quantities in the frozen regions.} In section \ref{sec:computation}, we describe various quantum extremal surface saddles. There we also discuss the entropy of a single inflating region. This leads us to a paradox which we discuss in section \ref{sec:paradox}. Then in section \ref{sec:resolution} we propose a resolution of this paradox via the gravitational path integral. In section \ref{sec:discussion} we end with some discussion and speculations about encoding a closed universe with inflating regions in a quantum system via the gravitational path integral.

\section{JT gravity coupled to conformal matter in dS$_2$}\label{sec:setup}
\begin{figure}
\hspace{-10mm}\includegraphics[scale = .2]{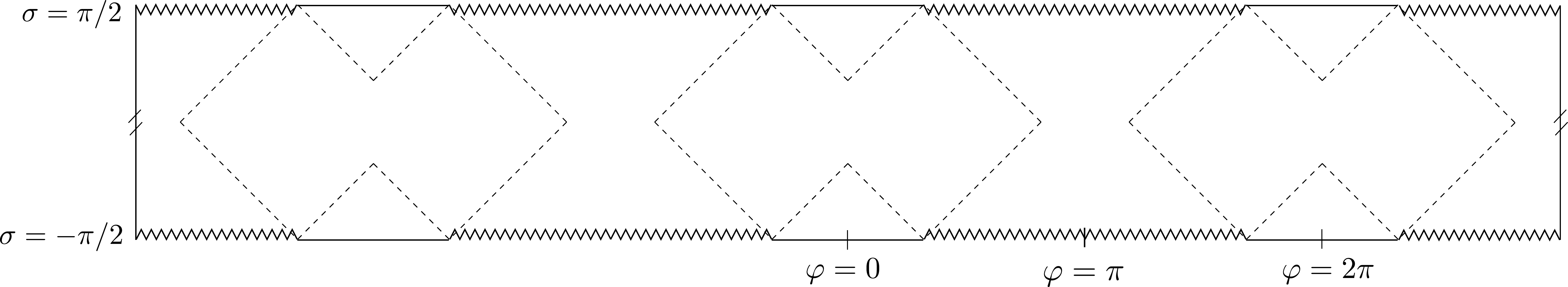}
\caption{Solution to JT gravity with $n=3$. The magnitude of the Casimir energy due to matter decreases as $n$ increases, leading to crunching regions which are larger than inflating regions. The Penrose diagram is periodically identified, making the spatial topology that of a circle.}\label{threeregions}
\end{figure}

We will consider Jackiw-Teitelboim gravity with positive cosmological constant minimally coupled to conformal matter: 

\be\label{JTact}
I = - \f{\phi_0}{4\pi}\left[\int_{\Sigma_2} R + 2 \int_{\p\Sigma_2} K \right] -\f{1}{4\pi} \left[\int_{\Sigma_2} \phi (R-2) + 2\phi_b \int_{\p \Sigma_2} (K-1) \right] + I_{\text{CFT}}\,.
\ee
 The path integral over the dilaton fixes us to dS$_2$,
\be
ds^2 = \f{ - d\s^2+d\varphi^2}{\cos^2 \s}\,,\qquad \varphi\sim \varphi + L \,,\qquad \s \in (-\pi /2, \pi /2).
\ee
We have fixed the de Sitter length to $1$. The metric equation of motion is 
\be\label{eom}
\left(g_{\m\n}\nabla^2 - \nabla_\m\nabla_\n+g_{\m\n}\right) \phi  = 2\pi T_{\m\n} \,.
\ee
The stress tensor on metric $ds^2 = e^{2\omega} ds^2_{\hat{g}} = e^{2\omega} (-d\s^2 + d\varphi^2)$ is given by
\begin{align}
T_{\m\n} &= T_{\m\n}^{\hat{g}} -\f{c}{12\pi}\left(\hat{\nabla}_\m \omega \hat{\nabla}_\n \omega - \f 1 2 \hat{g}_{\m\n} (\hat{\nabla}\omega)^2 - \hat{\nabla}_\n \hat{\nabla}_\m \omega + \hat{g}_{\m\n}\hat{\nabla}^2 \omega\right),\\
&= T_{\m\n}^{\hat{g}} +\f{c}{24\pi} \d_{\m\n} +\f{c}{24\pi} g_{\m\n}\,.\label{tfin}
\end{align}
In the final line, the second term is traceless and the last term is proportional to the metric. Picking the periodicity $\varphi \sim \varphi + 2\pi$ means the stress tensor on the cylinder $-d\s^2 + d\varphi^2$ is given by $T_{\m\n}^{\hat{g}} = -\f{c}{24\pi} \d_{\m\n}$, precisely canceling the piece $\f{c}{24\pi} \d_{\m\n}$ above. This leaves only the term proportional to the metric, which can be absorbed into a constant shift in $\phi$.\footnote{The only term in the equation of motion \eqref{eom} sensitive to constant shifts in $\phi$ is the $g_{\m\n}\phi$ term, which can therefore cancel a contribution to the stress tensor that is proportional to $g_{\m\n}$.} In this case we have a dilaton solution
\be
\phi = \phi_r \f{\cos \varphi}{\cos \s}.
\ee
We can consider a different periodicity for $\varphi$, $\varphi \sim \varphi + L$, in which case 
\be
T_{\m\n}^{\hat{g}} = -\f{\pi c}{6L^2} \d_{\m\n},\ \ \varphi \sim \varphi + L.
\ee
A $\varphi$-independent solution is given by
\be
\phi = -2\pi T_{\s\s}^{\text{traceless}}\left(1+ (\g+\s)\tan \s\right)\, ,
\ee
where $T_{\m\n}^{\text{traceless}}= T_{\mu \nu}^{\hat{g}} + \frac{c}{24\pi} \delta_{\mu \nu}$ is the traceless part of $T_{\mu \nu}$. Since this satisfies the sourced equation (i.e. $T_{\m\n}^{\text{traceless}} \neq 0$), we can add it to our previous sourceless solution to obtain
\be
\phi = \tilde{\phi}_r\f{\cos \varphi}{\cos \s}  -2\pi T_{\s\s}^{\text{traceless}}\left(1+ (\g+\s)\tan \s\right)
\ee 
for some free constant $\tilde{\phi}_r$ and where now $\varphi \sim \varphi + L$ for general $L$. We pick $L = 2\pi n $ with $n \in \mathbb{Z}^+$ to ensure periodicity of our dilaton $\phi$, obtaining
\be
\phi = \tilde{\phi}_r\f{\cos \varphi}{\cos \s}  -\f{c}{12}\left(1-\f{1}{n^2}\right)\left(1+ (\g+\s)\tan \s\right),
\ee
For a time-symmetric solution around $\s = 0$ we pick $\g = 0$, and to ensure we are inflating in at least some region of $\mathcal{I}^+$ we pick $\tilde{\phi}_r  > \f{\pi c}{24}(1-1/n^2)$. We also drop the constant piece in $\phi$, giving altogether
\be
\phi = \tilde{\phi}_r \frac{\cos \varphi}{\cos \sigma} - \frac{c}{12} \left(1- \frac{1}{n^2}\right)\sigma \tan \sigma.
\ee
Expanding around $\epsilon = \pi/2 - \s$ gives 
\be
\phi \approx \f{\tilde{\phi}_r \cos \varphi - \f{\pi c}{24}\left(1-\f{1}{n^2}\right)}{\epsilon}\,.
\ee
We will refer to a region where $\phi \rightarrow -\infty$ as a crunching region (i.e. a black hole interior) and $\phi \rightarrow +\infty$ as an inflating region. We therefore see that this family of sourced solutions shrinks the inflating regions and grows the crunching regions as compared to the unsourced solutions. The inflating regions remain out of causal contact from one another, i.e. the (black hole) wormhole grows and therefore remains nontraversable. This is reasonable since we are reducing the magnitude of the Casimir energy. The case of $n=3$ is shown in figure \ref{threeregions}. Extending the periodicity of the universe as a model was suggested in \cite{Hartman:2020khs} and studied in \cite{Aguilar-Gutierrez:2021bns}.

\subsection{Matter entropy}\label{matent}
We will also need the matter entropy on our $2\pi n$-sized universe. The quantum state of matter will be given by a Weyl transformation of the vacuum state on the flat cylinder of size $2\pi n$. We therefore write
\be
ds^2 = \f{dx d\bar{x}}{\Omega^2}\,,\qquad \Omega = \f{1}{2n} (x \bar{x})^{(1-n)/2} (1+ (x \bar{x})^n) \,,
\ee
with a map to the dS metric given by 
\be
x = e^{-i(\s-\varphi)/n}\,,\qquad \bar{x} = e^{-i(\s+\varphi)/n}\,.
\ee
This gives the CFT entropy as
\be\label{eqn:entropyindS}
S_{CFT} = \f c 6 \log \left(\f{(x_2-x_1)(\bar{x}_2-\bar{x}_1)}{\epsilon_{UV}^2\Omega(x_1)\Omega(x_2)} \right)= \f c 6 \log \left( 2n^2 \f{\cos\left( \f{\s_2-\s_1}{n} \right)- \cos\left( \f{\varphi_2 - \varphi_1}{n}\right)}{\epsilon_{UV}^2\cos \s_1 \cos \s_2}\right),
\ee
where $\epsilon_{UV}$ is an arbitrary cutoff to make the argument of the logs dimensionless. Notice that the Euclidean background is singular for $n>1$, since the angular coordinate of the sphere satisfies $\varphi \sim \varphi + 2\pi n$. The state of matter is still well-defined due to the Weyl equivalence with a smooth background. The inclusion of the gravitational sector breaks this equivalence, although to formulate our paradox we will assume there exists a reasonable state for the gravitational sector which allows us to use the island rule. We will return to this point in section \ref{sec:resolution}.

\subsection{Backreacted and extended Nariai solution}
Recall that JT gravity in dS$_2$ can be obtained by a dimensional reduction of near-extremal black holes in dS$_d$. These near-extremal black holes are the Schwarzschild black hole in the limit where the black hole horizon approaches the cosmic horizon, called the Nariai limit. In the higher-dimensional picture, this spacetime has an analytic extension which puts in as many inflating and crunching regions as desired. Figure \ref{threeregions} is simply a dimensional reduction of one of these possible analytic extensions, for the case $n=3$. Usually, the extension to additional inflating regions is considered a mathematical curiosity; what are all these other universes? 

A sharper objection is that of Kay and Wald \cite{Kay:1988mu}, which argued that there are no reasonable quantum states for quantum fields on the extended Schwarzschild-de Sitter spacetime, which respect the isometries of the spacetime. They proved this two different ways. The first is effectively the statement that Schwarzschild-de Sitter is out-of-equilibrium, since the black hole horizon and cosmic horizon have different temperatures. This disallows a standard Euclidean preparation of a state. This argument does not apply to the Nariai limit we are concerned with where the two horizons have the same temperature. The second proof uses monogamy of entanglement. If you line up several bifurcate horizons in a row, then a single diamond between the black hole and cosmic horizons has to purify both the diamond to its left and the diamond to its right in a state which respects the de Sitter symmetries. But this is impossible unless the left and right diamond are the same, as occurs in the $n=1$ spacetime. The way the state described in the previous subsection evades this argument is that the bifurcate horizons disappear once we consider quantum corrections to the spacetime solution. This suggests that the quantum state for matter we discussed above -- and the analytically extended spacetime -- may be fact instead of fiction.

It would be nice to study the same issue in higher dimensions. A time slice of the Nariai geometry is $S^1 \times S^2$, and for thermal periodicity conditions along the spatial $S^1$ we expect a negative Casimir energy, whose magnitude decreases as we grow the size of the $S^1$. This provides a contribution which wants to make the black hole wormhole grow as in two dimensions. We derive monotonicity of the Casimir energy with the length of the circle $L$ and some further constraints for a conformal field theory on $S^1 \times S^2$ in appendix \ref{app:casimir}. (The Nariai geometry is time-dependent so here we are talking about the instantaneous energy, say at $\s = 0$.)  

We need not take the other universes in the analytic extension seriously as a phenomenological model for what happens in our universe. Indeed, black holes formed from collapse do not look like this. But similar to the thermofield double in anti-de Sitter space, it is a useful theoretical model to probe various questions about horizons. 

\subsection{Boundary conditions}
\begin{figure}\centering
\hspace{-10mm}\includegraphics[scale = .3]{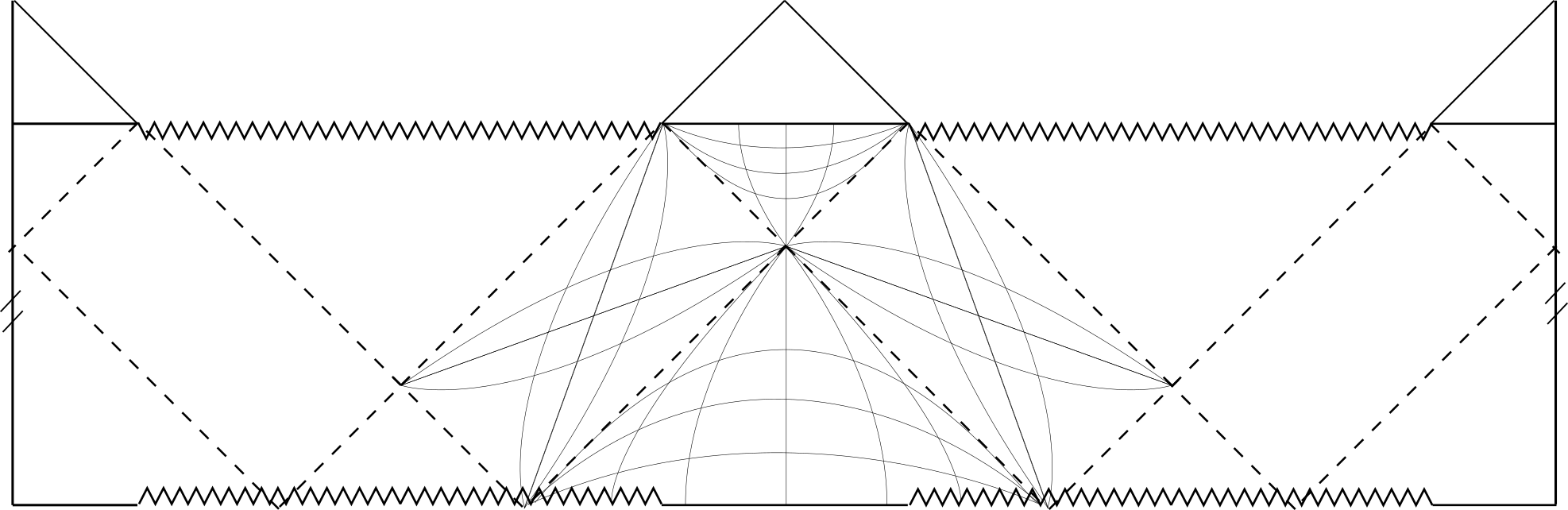}
\caption{The lines in this diagram illustrate constant $\tilde{\eta}$ and $\tilde{\chi}$ surfaces. }\label{newcoords}
\end{figure}

If we want to involve our saddle in a Hartle-Hawking-like path integral prescription, we need to know what boundary conditions to put at the future boundary. We will cut-off the space-time and glue to flat-space at the location defined by
\begin{align}
    \phi(x) = \frac{\phi_r}{\epsilon}\,,
\end{align}
and we will fix the induced metric on this curve to be
\begin{align}
    ds^2 = \frac{dx^2}{\epsilon^2}\,.
\end{align}
The flat-space metric will be
\begin{align}\label{eqn:hatmetric}
    ds^2_{hat} = \frac{-dt^2 + dx^2}{\epsilon^2}\,.
\end{align}
We will often refer to this flat-space region as a \emph{hat}, represented by the triangles at the top of figure \ref{newcoords}. If this gluing occurs near $\mathcal{I}^+$, then the boundary condition on the dilaton picks out a curve $\sigma(\varphi)\approx \f{\pi}{2} - \d \s(\varphi)$ in global coordinates which obeys the equation 
\begin{align}
    \tp \frac{ \cos \varphi}{\delta \sigma(\varphi)} - \f{c}{12}\left(1-\f{1}{n^2}\right)\frac{\frac{\pi}{2} -\delta \sigma(\varphi)}{\delta \sigma(\varphi)} = \frac{\phi_r}{\epsilon}. 
\end{align}
Solving for $\delta \sigma(\varphi)$ we have
\begin{align}\label{eqn:cutoffglobal}
    \delta \sigma(\varphi) = \epsilon\, \frac{\tp}{\phi_r} \left( \cos(\varphi) - \alpha_n\right),\ \ \alpha_n \equiv \frac{c\pi(1-1/n^2)}{24\tp}.
\end{align}
As of now, the ratio $\tp/\phi_r$ is an undetermined constant (analogous to $\eta_c/\epsilon$ in \cite{Chen:2020tes}) which in principle will need to be fixed in some auxiliary manner. We will return to this point shortly. Note that $\delta \sigma(\varphi)$ goes to zero when $\cos \varphi_* = \alpha_n>0$. This means that the inflating region goes from $\varphi \in (-\arccos \alpha_n, \arccos \alpha_n)$ with $\arccos \alpha_n <\pi/2$. In other words, the backreaction of the quantum fields on the $n>1$ universe causes the inflating region to shrink and the wormhole to grow, as was discussed above.

There is a natural Milne-like wedge which covers the causal past of the portion of $\mathcal{I}^+$ that is inflating. Focusing on the inflating region which is centered about $\varphi = 0$, we can define coordinates which cover this wedge by the equations (see Appendix \ref{app:coords})
\begin{align}\label{eqn:milnecoords}
    &\tanh(\tc) = \sqrt{1-\alpha_n^2} \frac{\sin \varphi}{\sin \sigma - \alpha_n \cos \varphi} \nonumber \\
    & \tanh(\te) =\sqrt{1-\alpha_n^2} \frac{\cos \sigma}{\alpha_n \sin \sigma -\cos \varphi}.
\end{align}
One can also check that the dS$_2$ metric in these coordinates is just the familiar de-Sitter metric in the Milne wedge
\begin{align}\label{eqn:milnemetric}
    ds^2 = \frac{-d\te^2 + d\tc^2}{\sinh^2 \te},
\end{align}
and furthermore note that the cutoff surface given by $\delta \sigma(\varphi)$ in \eqref{eqn:cutoffglobal} is at constant $\te$ given by 
\begin{align}
\tanh \te_c \approx \tilde{\eta}_c = -\epsilon\sqrt{1- \alpha_n^2} \frac{\tp}{\phi_r}.
\end{align}
We see that we can continuously match $\te$ and $\tc$ with the $t$ and $x$ coordinates of \eqref{eqn:hatmetric} to get the metric in the hat
\begin{align}\label{eqn:hatmilne}
    &ds_{hat}^2 = \frac{-d\te^2 + d\tilde{\chi}^2}{\tilde{\eta}_c^2}.
\end{align}
As discussed in \cite{Chen:2020tes}, the ratio $\tilde{\eta}_c/\epsilon$ (or $\tilde{\phi}_r/\phi_r$) has to do with the re-scaling between the flat space $x$ coordinate and the Milne $\tc$ coordinate. The parameterization of the global manifold in terms of these coordinates (and their analytic continuations) is presented in figure \ref{newcoords} for the case $n=2$.

To fix this undetermined parameter, one should compute the norm of the multi-hat state using path integral methods. If our saddle dominates this path integral, then the multi-hat state's norm will depend explicitly on $\te_c/\epsilon$. One can then extremize this norm over all possible values of $\te_c/\epsilon$. See \cite{Chen:2020tes} for further discussion of this point. As is discussed below in section \ref{sec:resolution}, the saddles we have discussed here do not naturally dominate the path integral. Without a specific prescription for making this multi-hat saddle dominant, we cannot explicitly fix the ratio $\te_c/\epsilon$. Thus, for the remainder of this work, we will leave it as an unfixed parameter, keeping in mind that in principle it will be fixed to a specific value.
We can use our coordinate transformation in \eqref{eqn:milnecoords} to write the metric \eqref{eqn:hatmilne} in terms of $\sigma, \varphi$ coordinates. It takes the form
\begin{align}
    &ds_{hat}^2 = \frac{\sinh^2 \te(\sigma,\varphi)}{\te_c^2 \cos^2 \sigma} \left(-d\sigma^2 + d\varphi^2\right) \nonumber \\
    & = \Omega^2(\sigma,\varphi) \left(-d\sigma^2 + d\varphi^2\right),
\end{align}
where 
\begin{align}\label{eqn:hatWeyl}
    \Omega^2(\sigma, \varphi) = \frac{1}{\te_c^2} \frac{1-\alpha_n^2}{(\cos \varphi - \alpha_n \sin \sigma)^2-(1-\alpha_n^2) \cos^2 \sigma}.
\end{align}
We can then find entanglement entropies for regions with one endpoint in the hat and the other in the de-Sitter region by simply replacing one of the $\Omega$'s in \eqref{eqn:entropyindS} by the $\Omega$ in \eqref{eqn:hatWeyl}. Since $\Omega$ is local to the endpoint in the hat region, this will just affect the answer for the entropy in \eqref{eqn:entropyindS} by an overall constant, independent of the position of the endpoint that is in the de-Sitter region.

Finally, an important but potentially confusing point is that the quantum fields of the CFT living in the back-reacted Milne wedge covered by the coordinates in \eqref{eqn:milnecoords} will \emph{not} be thermal with respect to Milne time $\te$. There is, however, a different set of coordinates which one can choose for the same back-reacted wedge with respect to which the CFT state is thermal. We can find these coordinates by first conformally mapping the interval $\sigma = \pi/2$, $\varphi \in (-\varphi_*, \varphi_*)$ to the half-circle $\sigma = \pi/2$, $\varphi \in (0,\pi n)$. The wedge associated to this half-circle can then be viewed as a Rindler wedge of the Poincare patch of the Lorentzian cylinder covered by $\sigma, \varphi$. Following this procedure, the Rindler coordinates covering this Rindler wedge are related to $\sigma, \varphi$ by
\begin{align}\label{thermcoords}
    &\tanh x_{\text{th}} = \sqrt{1-\beta_n^2} \frac{\sin \frac{\varphi}{n}}{\cos \frac{\sigma - \pi/2}{n} - \beta_n \cos \frac{\varphi}{n}}\nonumber \\
    &\tanh t_{\text{th}} = -\sqrt{1-\beta_n^2} \frac{\sin \frac{\sigma - \pi/2}{n}}{\beta_n \cos \frac{\sigma - \pi/2}{n} - \cos \frac{\varphi}{n}} 
\end{align}
where $\beta_n = \cos \frac{\varphi_*}{n}$ with $\varphi_*$ defined by $\alpha_n = \cos \varphi_*$. See Appendix \ref{app:coordstherm} for details. Just as before, one can check that the coordinates $t_{\text{th}}$ and $x_{\text{th}}$ cover the wedge associated to the central inflating region. To reiterate, the state of the fields \emph{is} thermal with respect to $t_{\text{th}}$ but \emph{not} $\te$.

\section{Island computation}\label{sec:computation}
With the gravitational solutions at hand, we want to consider the generalized entropy of an interval in one of the inflating regions, analogous to the computations in \cite{Chen:2020tes, Hartman:2020khs, Aguilar-Gutierrez:2021bns}. We will slightly modify the solution above by appending flat-space regions to each of the inflating regions. Since the dilaton diverges toward the inflating boundary, the gravitational coupling is approaching zero there. Therefore in the flat-space region we will assume gravitational effects can be completely ignored.

\begin{figure}
\hspace{-10mm}\includegraphics[scale = .2]{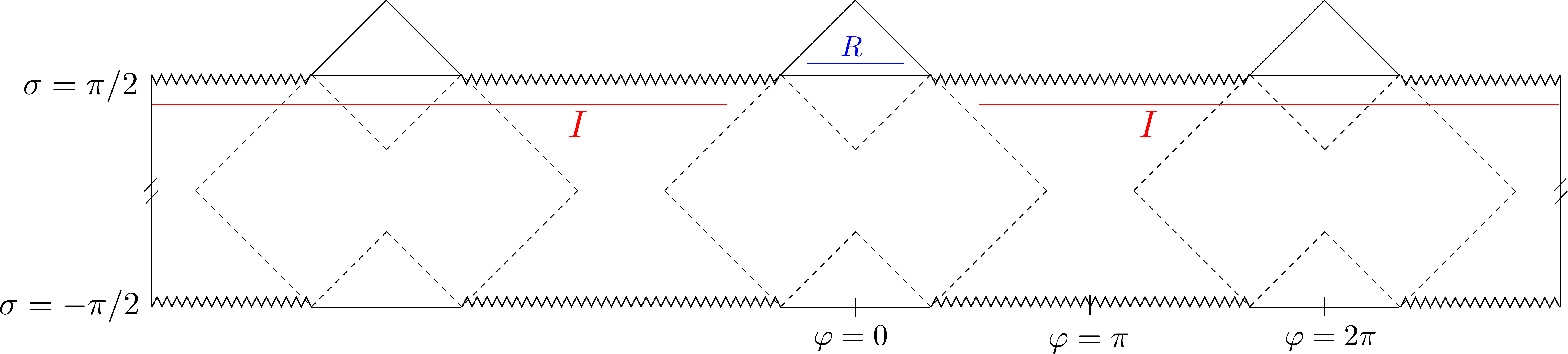}
\caption{Island region $I$ for entropy of region $R$ in an $n=3$ universe.  }\label{3island}
\end{figure}

We will assume that the island region is as depicted in figure \ref{3island}, and that we are in an OPE limit such that the entropy in the complement channel -- which is the union of two intervals -- factorizes. We will do the computation for a region $R$ close to but below $\mathcal{I}^+$, where we will ignore the effects of gravity; as discussed below \eqref{eqn:hatWeyl}, moving $R$ into the hat as in figure \ref{3island} simply introduces an additive constant factor due to the distinct Weyl factor in the hat. The gravitational entropy is $\phi/4$, which when combined with the matter entropy in section \ref{matent} gives the generalized entropy as:
\be
S_{gen} = 2\phi_0 +2 \tilde{\phi}_r \f{\cos \varphi_I}{\cos\s_I} - \f{c}{6}\left(1-\f{1}{n^2}\right) \s_I \tan \s_I+\f c 3 \log \left(2n^2 \f{\cos\left( \f{\s_I-\s_R}{n} \right)- \cos\left( \f{\varphi_I - \varphi_R}{n}\right)}{\epsilon^2\cos \s_I \cos \s_R}\right)
\ee
The overall factor of two is for both intervals. We want to extremize this answer with respect to the $\{\s_I, \varphi_I\}$ endpoint. It is a little difficult to extremize this directly, but in the limit of small backreaction $c/\tilde{\phi}_r \ll 1$,  
the $n>1$ saddles are not very different from the $n=1$ saddle, at least as long as we choose the endpoints of region $R$ to be near the interfaces between the crunching and inflating regions on $\mathcal{I}^+$.
The resulting island is as in figure \ref{3island}.

The above situation presents us with a puzzle. While an observer Alice with access to region $R$ can encode the rest of the universe, the same would apply to an observer Bob in the right or left patches. In particular, Alice and Bob would have overlapping islands (and would be in each other's island). This is inconsistent with complementary recovery, and leads to a violation of the no-cloning theorem in quantum mechanics. 

\subsection{Entropy of an entire hat}

\begin{figure}
\centering
\includegraphics[scale = .2]{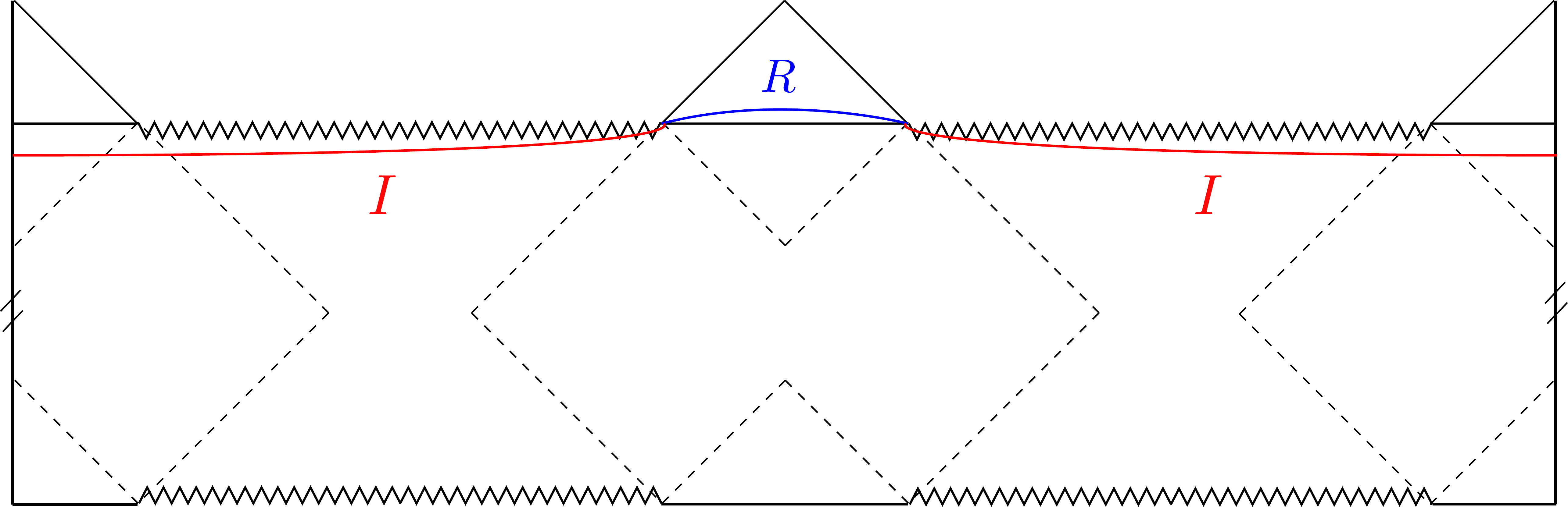}
\caption{The ``purity" saddle which gives $S = 0$ for the entropy of one of the two hats in an $n=2$ universe.}\label{2hats}
\end{figure}

We can also take region $R$ to be an entire inflating region. In this case, the natural answer for the entropy, analogous to the island we found in the previous section, is to extend region $R$ into the bulk such that it covers the entire spacetime. This is displayed in figure \ref{2hats} and gives $S = 0$. Like in the previous section, if we compute the entropy of the other hat, we will find again that the island region is the rest of the spacetime, giving $S = 0$ again. This can also be seen more directly from the replica analysis, which we discuss in section \ref{sec:resolution}. Now that we have argued for overlapping island regions, let's move onto the paradox that arises. 
\clearpage
\vspace{-3mm}
\vspace{-13mm}
\section{A paradox} \label{sec:paradox}
In this section we will carefully state our assumptions and the inconsistency they lead to. We will see that the following two assumptions are in contradiction with each other:
\begin{enumerate}
\item For the two (or multi) hat state, the entanglement wedge of either hat is the entire universe.
\item Any operator in Hat$_1$ commutes with any operator in Hat$_2$. 
\end{enumerate}
We now argue by contradiction that these both cannot be true. By the first assumption, the entanglement wedges of the two hats overlap, for example in either of the black hole interiors. For operators $\phi$ and $\pi$ in the black hole interior such that $[\phi, \pi] \neq 0$, we then have
\be
\langle \psi|[\phi, \pi] |\psi\rangle =\langle \psi| [O_1, O_2] |\psi \rangle = 0
\ee
where in the first equality we used entanglement wedge reconstruction to represent $\phi$ in Hat$_1$ with $O_1$ and $\pi$ in Hat$_2$ with $O_2$. The second equality follows by assumption 2, but then we reach a contradiction since we assumed $[\phi, \pi] \neq 0$. 

\subsection*{Connection to no cloning}
Note that this contradiction is very similar to the contradiction that if we have two overlapping entanglement wedges for complementary regions then we could clone quantum information. The proof for this is as follows. Suppose we have a quantum error correcting code with overlapping entanglement wedges for complementary regions. Denote the two complementary regions by $A$ and $\bar{A}$. Suppose our code-subspace $\mathcal{H}_C$ is spanned by the states $\lbrace \ket{i} \rbrace$ indexed by $i$. This subspace could be, for example, the Hilbert space of a qubit in the past of Hat$_1$ or Hat$_2$.

Then suppose that both of these regions can reconstruct the code-subspace. Reconstructability is equivalent to the existence of a decoding isometry which isolates the code subspace state onto a sub-factor of the physical Hilbert space, $\mathcal{H}_{A\bar{A}}$, with dimension equal to that of the code subspace. In other words, using the conventions and notations of \cite{Harlow_2017}, this means that there is an isometry $U_A$ acting on $\mathcal{H}_A$ such that 
\begin{align}\label{eqn:decode}
U_A \ket{i}_{A\bar{A}} = \ket{i}_{A_1} \otimes \ket{\chi}_{A_2 \bar{A}}
\end{align}
for all $\ket{i}$ in the code subspace and for some division $A_1$ and $A_2$ such that $|A_1| = \dim \mathcal{H}_C$ and $|A_2| = \dim \mathcal{H}_A/\dim \mathcal{H}_C$. See \cite{Harlow_2017} for the slight modification of this formula if $\dim \mathcal{H}_C$ is not a divisor of $\dim \mathcal{H}_A$, although this is unimportant for us. Here $\ket{\chi}_{A_2\bar{A}}$ is some state which is independent of $\ket{i}$, which is essential.

Now by assumption there is also a similar equality for $U_{\bar{A}}$ on the complement region. Putting eq. \eqref{eqn:decode} together with its complementary version, and using that $U_A$ and $U_{\bar{A}}$ commute, we have
\begin{align}\label{eqn:paradox}
U_{\bar{A}} U_A\ket{i}_{A\bar{A}} = \ket{i}_{A_1} \otimes U_{\bar{A}} \ket{\chi}_{A_2 \bar{A}} = \ket{i}_{\bar{A}_1} \otimes U_{A} \ket{\chi}_{\bar{A_2} A}.
\end{align}
The latter equality tells us that $\chi$ is in fact dependent on $i$, violating the assumption of reconstructability. Note that if this string of equalities were true then we could clone quantum information. This is because the second equality in eq. \eqref{eqn:paradox} tells us that the reduced density matrix of $U_{\bar{A}} U_A \ket{i}_{A\bar{A}}$ on $A$ is $\ket{i}\bra{i}_{A_1} \otimes \chi_{A_2}$, where $\chi_{A_2} = \text{Tr}_{\bar{A}} \ket{\chi} \bra{\chi}_{A_2 \bar{A}}$, and analogously for $\bar{A}$, but the only pure state on $A\bar{A}$ with these reduced density matrices is
\begin{align}
    U_{A}U_{\bar{A}} \ket{i}_{A\bar{A}} = \ket{i}_{A_1} \otimes \ket{i}_{\bar{A}_1} \otimes \ket{\chi}_{A_2 \bar{A}_2}.
\end{align}
Thus, the joint isometry $U_AU_{\bar{A}}$ would allow one to clone quantum information onto the $\mathcal{H}_{A_1} \otimes \mathcal{H}_{\bar{A}_1}$ subfactor of $\mathcal{H}_A$, which of course is impossible.

\section{Resolution}\label{sec:resolution}
Our proposed resolution to the paradox above is that the quantum extremal surface saddles in figures \ref{3island} and \ref{2hats} are actually \emph{incorrect} for the problem as posed. The reason will be roughly due to a ``time-like" homology constraint.  To illustrate this, we first examine a slightly different set-up than the one we consider with multiple disconnected hats. 
\begin{figure}
\centering
\includegraphics[scale = 1.2]{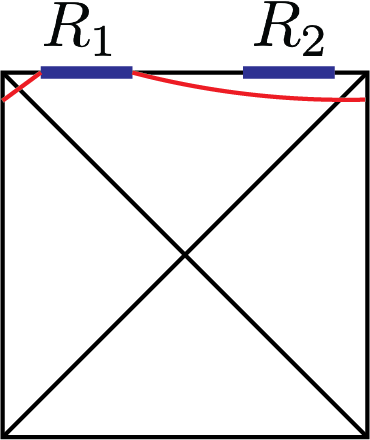}
\caption{We imagine taking global de Sitter and freezing two regions near $\mathcal{I}^+$ pictured in red. We integrate over the geometry away from these intervals. Naively, there is a puzzle since the entanglement wedge for one of the intervals appears to encompass the whole universe, since the spatial cross sections are compact. The naive entanglement wedge for $R_2$ is pictured in blue. }\label{fig:twoint}
\end{figure}
\subsection{A toy model of a toy model}\label{toytoy}
We briefly discuss a slightly simpler set-up where a similar confusion arises. Consider global $dS_{d+1}$, without black holes, illustrated in figure \ref{fig:twoint}. We can imagine ``freezing" the geometry in two regions $R_1$ and $R_2$ close to $\mathcal{I}^+$. By freezing here, we mean that in defining the quantum state near $\mathcal{I}^+$ we only integrate over quantum fields while fixing the metric in the frozen regions.\footnote{One might have in mind here that the two frozen regions $R_1$ and $R_2$ correspond to two boundary quantum field theories. The state of the system on $R_1 \cup R_2$ is then prepared via the path integral over geometries in its past. To determine this state, one could use the Hartle-Hawking prescription or perhaps a modified prescription to produce a different state, as we will discuss below.} 

If we take the saddle-point in figure \ref{fig:twoint} seriously, one would run into a paradox similar to the one described in the section \ref{sec:paradox}. This is because again the spatial cross-sections of global $dS_{d+1}$ are just topologically $S^d$, and so the entanglement wedge for either $R_1$ or $R_2$ is just determined by the trivial quantum extremal surface, i.e. the entanglement wedge is the whole universe. If this were true, there would be operators encoded in $R_1$ which do not commute with operators in $R_2$.

\begin{figure}
\centering
\includegraphics[scale = 1.3]{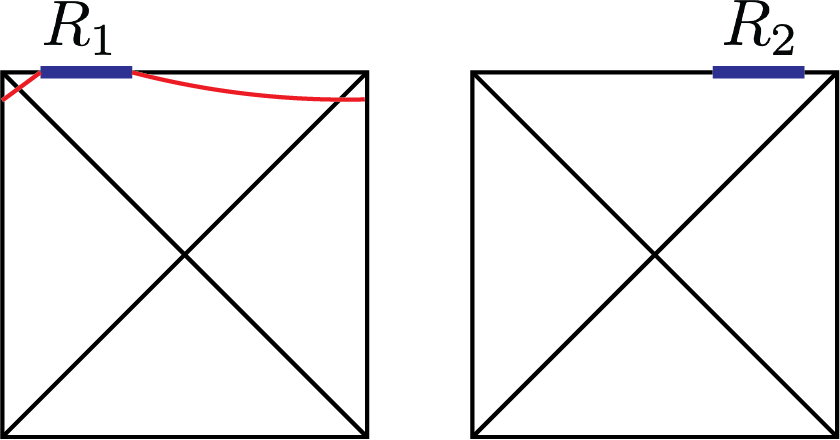}
\caption{The true saddle for the problem of two regions near $\mathcal{I}_+$ is actually two disconnected universes, with each region in its own copy of the original spacetime.}\label{fig:twointdisconn}
\end{figure}

We can see how the Hartle-Hawking prescription resolves this confusion, however. The HH prescription says that to compute the dominant contribution to the wave-function we just fix the two regions and then sum over all no-boundary geometries in the past which end on these intervals.\footnote{Note that, as always, there are contributions from closed universes which contain neither $R_1$ nor $R_2$. These are only relevant for computing the norm of the state but will divide out when we compute normalized quantities.} When we do this, however, the dominant saddle is \emph{not} the one pictured in figure \ref{fig:twoint}, but rather one where the two regions are in their own separate universes as in figure \ref{fig:twointdisconn}. This is because this saddle is enhanced by a factor of $e^{\frac{1}{16 \pi G_N} \int d^dx R} \equiv e^{S_0}$ (due to the additional universe) relative to the saddle where both intervals are in the same universe. In this saddle, there is no confusion: the two regions just encode their own copies of the universe. We see that without modification, the HH prescription produces a state of the two regions which is naturally disentangled.

One could ask if there is a modification of the HH prescription - in other words, a different state - where the saddle in figure \ref{fig:twoint} \emph{is} the dominant saddle. Instead of delving into this question more here, we instead turn to the main set-up of interest.

\subsection{Back to the multiple black hole set-up}\label{sec:multihatdom}

\begin{figure}
\centering
\includegraphics[scale = .8]{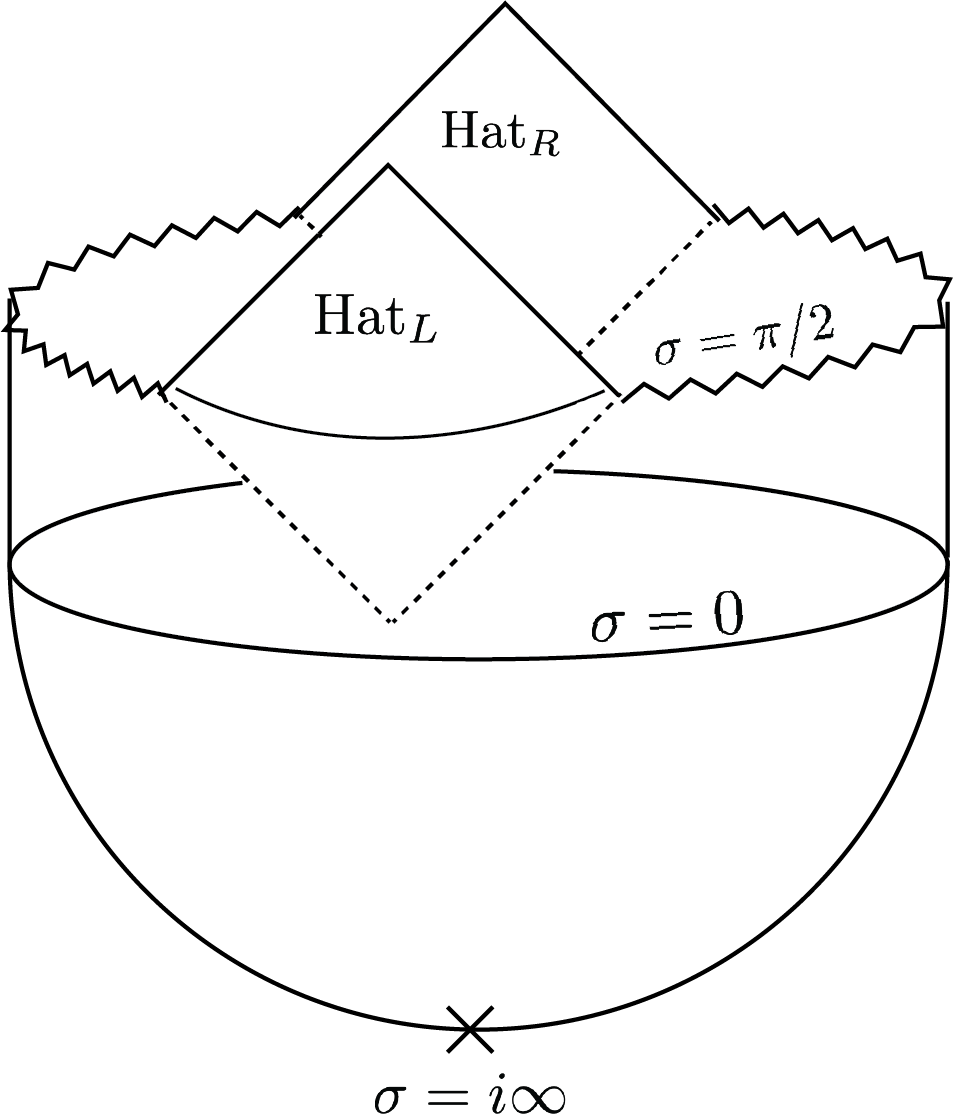}
\caption{The connected, two-hat saddle which we are interested in studying. Here we have illustrated the Lorentzian to Euclidean continuation of this saddle from the $\sigma = 0$ line. This is a hemisphere of curvature $R=2$ but with a conical excess at $\sigma = i\infty$, the south pole of the hemisphere. The conical singularity has an opening angle of $4\pi n$. For $n$ hats, it would have an opening angle of $2\pi n$.}\label{fig:twohat3d}
\end{figure}
We return to our model of 2d Schwarzschild-de Sitter with a $2\pi n$-sized universe for $n>1$. As mentioned at the end of section \ref{matent}, the Euclidean manifold which would prepare the Hartle-Hawking state has a conical singularity. In other words, although the configuration in Section \ref{sec:setup} is a solution everywhere in Lorentzian signature, the analytic continuation of these geometries to Euclidean signature is not everywhere a solution to the JT saddle-point equations. The spatial cross section of the multi-hat saddle depicted in figure \ref{2hats} has a total angle of $2n\pi$ for $n$ hats, which when continued into Euclidean signature leads to a conical singularity in the past ($\sigma = +i\infty$ in global coordinates), at which point the constraint $R=2$ is no longer obeyed.
This is illustrated in figure \ref{fig:twohat3d}. Say we have a UV complete theory where this conical singularity is regularized somehow. Then using the argument in Section \ref{toytoy}, we see that if we freeze $m < n$ hats, then the dominant saddle will be $m$ disconnected universes, and we will not run into a paradox of overlapping entanglement wedges. 

Without such a UV complete theory, the resolution is even simpler: these problematic spacetimes are simply not saddles. They instead appear as though an operator has been inserted at some point in the Euclidean past. Thus they do not contribute to -- let alone dominate -- the Hartle-Hawking path integral without operator insertions. 
\begin{figure}
\centering
\includegraphics[scale = .8]{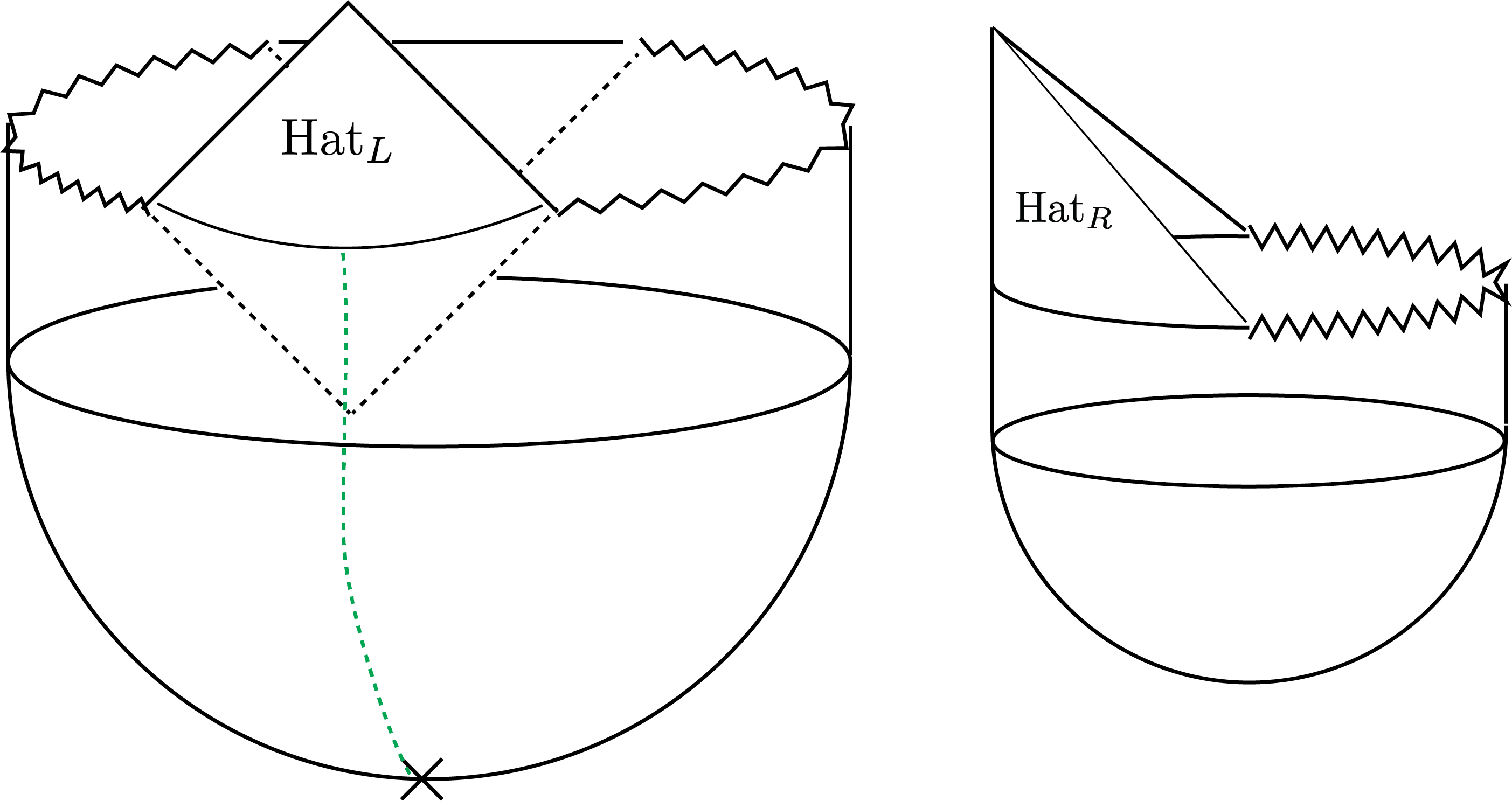}
\caption{If we insert a conical singularity at a position which references only one of the hats, instead of both simultaneously, the dominant saddle will be two disconnected universes illustrated here. If we define the conical singularity with respect to the Hat$_L$, then Hat$_L$ will be in a universe with angle $4\pi$ and another, fluctuating asymptotically inflating region. Hat$_R$ will be off in its own $2\pi$ universe. We have schematically illustrated the procedure of tying the conical singularity to a hat by the green dashed line. To prevent the saddle illustrated here from dominating the path integral, we need to tie the position of the conical singularity to both hats simultaneously. We discuss some ways of doing this in the main text.}\label{fig:3ddisconn}
\end{figure}

One could then ask: can we include operator insertions such that the connected spacetime becomes a solution, and in fact dominates the path integral? For this to occur, we need a nonperturbative definition of the location of the insertion, i.e. the location of the conical singularity. A natural way of discussing the location is to define it relative to the future boundary conditions. For example, one might geodesically ``dress" this point to the future asymptotic boundary. Then it is not hard to see that to make the connected two-hat saddle dominant, we need that the position of this singularity is defined relative to not just one of the hats \emph{but to both simultaneously.} To understand this, imagine that we referenced just one of the hats, say the left hat Hat$_L$. Then, similarly to the previous subsection, the dominant saddle will again be one where the two hats sit in their own, disconnected de Sitter universes since this is enhanced relative to the connected saddle by factors of $e^{S_0}$. Here the universe with Hat$_L$ has an extra asymptotically de Sitter region due to the conical singularity in the Euclidean region of the manifold. This is illustrated in figure \ref{fig:3ddisconn}. This again shows that Hartle-Hawking-like prescriptions naturally want to disconnect all asymptotically de Sitter regions, unless we force them to connect. Thus we see that to force them to connect we need to define the position of the conical singularity with respect to \emph{both} hats simultaneously. In this case the disconnected saddles no longer contribute since the dressing of the conical singularity only includes geometries which are connected in the path integral.

\subsection{Inserting the conical singularity relative to both hats and resolution of the paradox}
This discussion so far has been abstract since we have not discussed any concrete methods by which to actually insert this conical singularity in the Euclidean past. We now discuss two possible options.

\subsubsection{Freezing by hand}\label{sec:frozen}
\begin{figure}
\centering
\includegraphics[scale = .79]{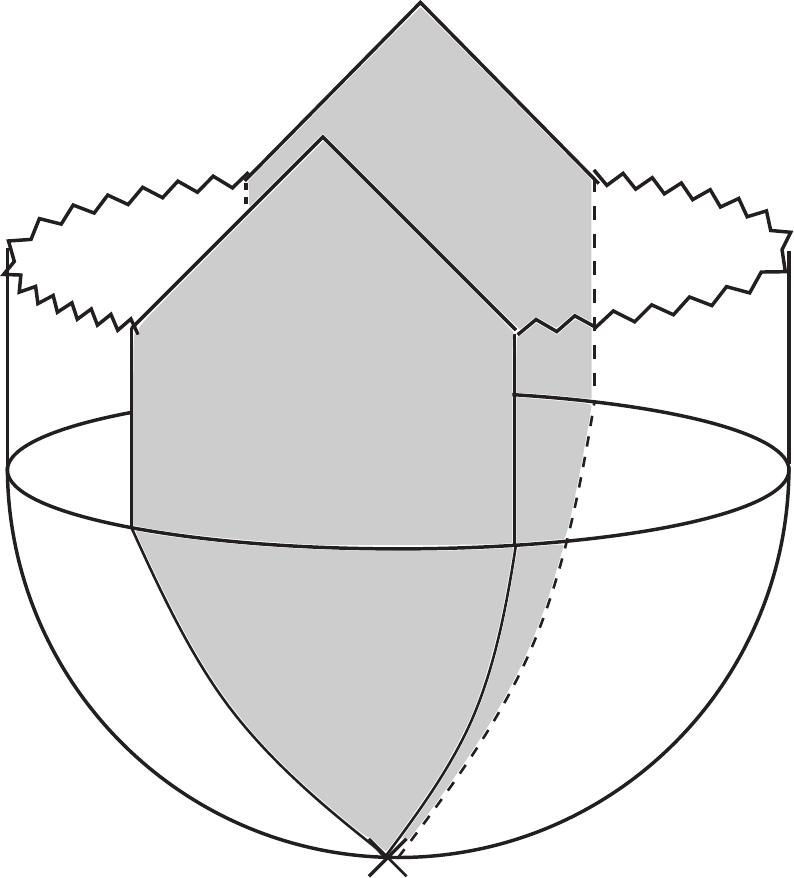}
\caption{One could make the two-hat connected saddle dominate the path integral by freezing some portion of the geometry in the interior. A natural choice is to freeze all regions with some dilaton value $\phi \geq 0$. Here we show the saddle where we have frozen everything filled in with gray. Note that when we include $c/\tilde{\phi}_r$ corrections to this solution then the frozen region actually includes a neighborhood of the conical singularity.}\label{fig:twohat3dfrozen}
\end{figure}

One option is to follow in the footsteps of \cite{Almheiri_2021} and just freeze more of the geometry by hand. For example, instead of just freezing in the asymptotic regions where the dilaton is becoming large and so gravity is becoming weak, we could for example choose to freeze all regions of the geometry where the dilaton $\phi(x)$ is bigger than some value $\phi_*$. For example, one could choose to freeze all parts of the geometry (Lorentzian or Euclidean) where the dilaton is $\phi(x) \geq 0$, which corresponds to all regions where the total dilaton is greater than its extremal value $\phi_0$. Note that this is not the value of the dilaton at the de Sitter horizon, which is instead $\phi(x) = \tilde{\phi}_r(1+ \mathcal{O}(c/\tilde{\phi}_r))$. 
Ignoring the quantum corrections of order $c/\tilde{\phi}_r$, we see that this amounts to freezing everything in the geometry with angular variable $\varphi \in [-\pi/2, \pi/2] \cup [3\pi/2, 5\pi/2]$. This is illustrated in figure \ref{fig:twohat3dfrozen}. We see that the conical singularity is just barely included in the Euclidean past frozen region.

With quantum corrections, the dilaton in fact linearly blows up to $+\infty$ at the conical singularity and so the frozen region, where $\phi \geq 0$, includes a neighborhood of the conical singularity. Since the geometry is frozen in what used to be the ``bulk" of the spacetime, we are considering a genuinely \emph{different} state from the HH state on the Hilbert space $\mathcal{H} = \mathcal{H}_{\text{Hat}_L} \otimes \mathcal{H}_{\text{Hat}_R}$. Given this frozen region, we then compute the wavefunction of this state by summing over all geometries which end on this mixed-signature manifold. Clearly, our connected saddle geometry is a solution to this problem and likely nothing else is. 

\textbf{Resolution of the paradox for frozen geometries}: If we freeze portions of the geometry by hand, then the resolution to our paradox is simple; since the frozen region extends into the Lorentzian past of both saddles, the naive QES for Hat$_L$, which includes the whole universe, can no longer work because it would manifestly include a piece of the frozen region in the past of Hat$_R$.

We can be more explicit and review how the Euclidean path integral implements the constraint that the entanglement wedge for a frozen region $R$ should not include any other frozen region besides $R$. To illustrate how the gravity path integral enforces this constraint, we can compute the Renyi-2 entropy, $S_2 = -\log \text{Tr} \rho^2$, of Hat$_L$ in the HH state prepared by JT gravity along with the extra boundary condition of freezing the portions of its geometry in figure \ref{fig:twohat3dfrozen}. We will call the state of the two hats prepared via this path integral $\ket{HH_f}$ where the subscript $f$, for ``frozen,'' denotes that we are working with the extra boundary conditions. What we want to argue is that $S_2 >0$ since the saddle which gives $S_2 = 0$ does not obey the boundary conditions.

\begin{figure}
\centering
\includegraphics[scale = 1]{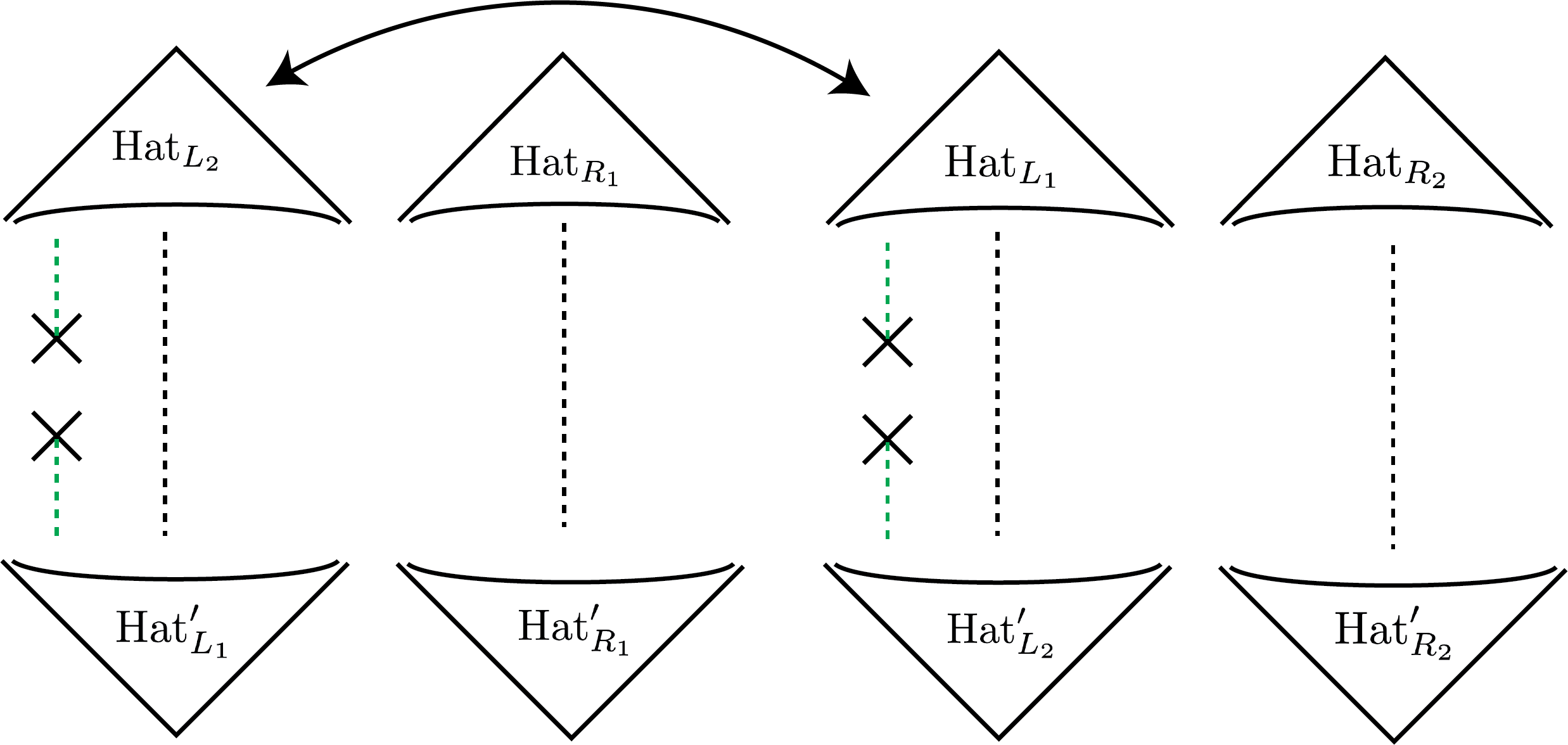}
\caption{This figure illustrates what happens if the frozen portion of the geometry is only connected to one hat, in this case Hat$_L$. The green-dashed lines schematically represent the frozen portion of the geometry. In this case, Hat$_L$ from replica 1 can be swapped with Hat$_L$ from replica 2 and so the dominant saddle will just be two copies of the saddle which dominates the norm, $\braket{HH_f|HH_f}$. In other words, $\text{Tr}[\rho^2] =\braket{HH_f|HH_f}^2$ where $\rho$ is the unnormalized density matrix.  The double-headed arrow indicates that the left hats from each replica have been swapped.}\label{fig:swaptrick}
\end{figure}
\begin{figure}
\centering
\includegraphics[scale = 1]{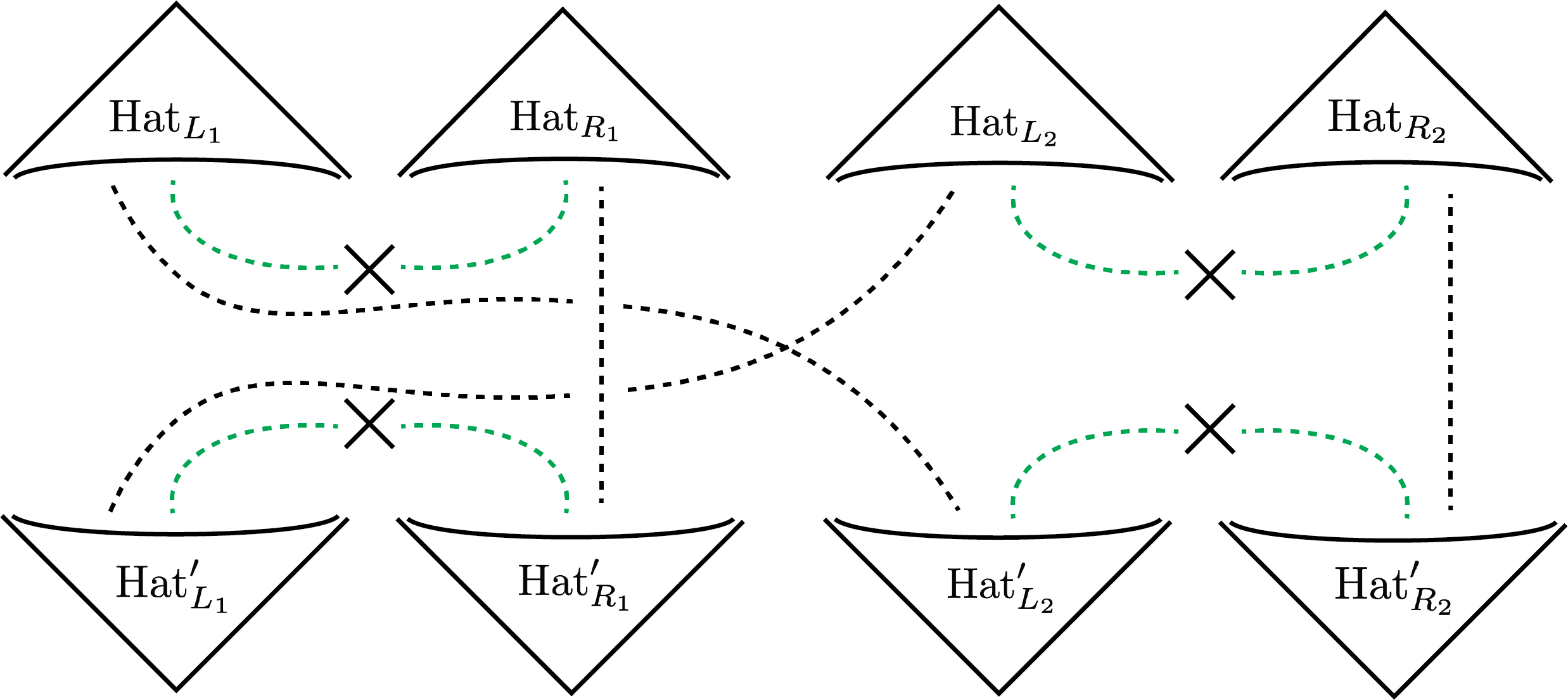}
\caption{We illustrate the boundary conditions associated to computing the purity, $\text{Tr}[\rho^2_L]$, for the state on Hat$_L$ in the Hartle-Hawking state with modified boundary conditions, where the conical singularity (marked by ``X") is inserted in the frozen portion of the geometry. The prime denotes bra vs. ket and the subscript denotes replica number. Since this dressing ties hats of the same replica number to each other, we see that the quantum extremal surface (or its Renyi-2 analog), which gave an answer of $S_2 = 0$ for each hat, is excluded. The only possibility is the analog of the ``Hawking" saddle, which gives a non-zero answer for $S_2$.}\label{fig:purityreptrick}
\end{figure}

To compute $S_2$, we prepare two replica copies of the pure state density matrix $\ket{HH_f}\bra{HH_f} \equiv \rho_{LR}^{HH_f}$, trace out $R$ and then compute $\text{Tr}[\rho_L^2]$. This amounts to computing the path integral with 4 boundaries or 8 asymptotic hat regions, gluing all the $R$ kets to their bra partners in the same replica and then gluing the $L$ kets to the $L$ bras in the other replica. This is illustrated in figure \ref{fig:purityreptrick}. 
To compute the purity, we are then instructed to sum over all JT saddles that have these boundary conditions discussed in the previous section. If the frozen region does not geometrically connect Hat$_L$ to Hat$_R$, then we can effectively swap the bra for Hat$_L$ in replica copy 1 with the bra in Hat$_L$ for replica 2 as in figure \ref{fig:swaptrick}. 
The dominant saddle would then just be two copies of the saddle that dominates when computing $\braket{HH_f|HH_f}$. In other words, we would get $S_2 = 0$.

If on the other hand the frozen region connects both boundaries as in figure \ref{fig:twohat3dfrozen}, then we cannot swap the replica copies because Hat$_L$ and Hat$_R$ are effectively tied to each other within each replica byt the path integral over the forzen region. In other words, one can only get saddles which connect between bras (or kets) of the same replica. This is not the same as $\braket{HH_f|HH_f}^2$ and so $S_2 \neq 0$. Indeed, we expect it to be of order $\phi_0$. As we will see in the next section, there is another quantum extremal surface which we have thus far ignored that appears for $n >1$ and which gives an answer $S_2 \approx \phi_0$.

\subsubsection{Modifying the JT action}
Freezing a large portion of the interior geometry is a rather drastic way of making the multi-hat saddle dominate the path integral. One might hope that there is a less severe way of accomplishing this goal. One method might be to modify the bulk theory so as to insert a conical singularity at the correct point. More explicitly, one could attempt to produce this saddle by modifying the JT action from 
\begin{align}\label{eqref:modJT}
\int d^2 x \phi(R-2) \to \int d^2 x \sqrt{-g} \phi \left( R-2-\frac{\alpha}{\sqrt{-g}} \delta^2(x-x_*) \right)
\end{align}
where $\alpha$ is an independent parameter that can be tuned to be $\alpha = \frac{2(n-1)}{n}$ with $n$ the number of asymptotic hats. Here we have in mind that $x_*$ is defined in some way that makes sense non-perturbatively, and, when evaluated on the metric $g_0$ with a conical singularity at the past Euclidean south pole (i.e. $\sigma = i \infty$ in global coordinates), we find that $x_*^{\mu}(g_0) = (\sigma = i\infty, \varphi )$. For more general metrics, $x_*$ might depend on the background metric and dilaton: $x_* = x_*(\phi, g)$. Suppose there exists some method to specify $x_*$ that depends only upon the background metric $g$ and not the dilaton. For example, we might imagine geodesically dressing the point $x_*$ to the future asymptotic boundary. We can then integrate over the dilaton and find that the geometries localize to those with positive curvature and a delta function singularity at $x = x_*$. The metric equations also get modified to be 
\begin{align}\label{eqn:dileqnmod}
\left(g_{\m\n}\nabla^2 - \nabla_\m\nabla_\n+g_{\m\n}\right) \phi  = 2\pi T_{\m\n} +\alpha \frac{\delta x^{\beta}_*(g)}{\delta g^{\m \n}(x)} \partial_{\beta} \phi(x_*)\,.
\end{align}

Note that for a point $x_*$ which is extremal with respect to the dilaton, this extra source term on the right hand side vanishes. This is why the fixed area states of \cite{Akers:2018aa, Dong:2018aa} have been considered for extremal surfaces only. More generally, however, this term may not vanish. Furthermore, depending on how specifically $x_*$ is defined in terms of $g$, $\delta x_*/\delta g$ might be quite hard to calculate. Regardless of the specific details of the dilaton solution to eq. \eqref{eqn:dileqnmod}, as long as there remain $n$ asymptotically inflating regions after accounting for the term proportional to $\alpha$, one can still discuss a version of our paradox in this modified saddle point. This paradox for the modified saddle described by \eqref{eqn:dileqnmod} will be resolved in the same way as follows. 

Figures \ref{fig:swaptrick} and \ref{fig:purityreptrick} can be used again in this context just by re-interpreting the green dashed lines as schematically denoting the dressing of the conical singularity with respect to the asymptotic hats, i.e. a prescription for defining $x_*(g)$. The same words then go through. When we dress the conical singularity to only one hat, the dominant saddle is the disconnected one. When we dress the conical singularity to both simultaneously, we can no longer swap hats in different replicas, since they are ``tied together'' by the green lines in figure \ref{fig:purityreptrick}. We see that the resolution is effectively the same, regardless of the details of how we prepared the entangled state of the two hats via the gravitational path integral.

\subsection{Entropy of an entire hat revisited}\label{hatentropy}
\begin{figure}
\centering
\includegraphics[scale = .2]{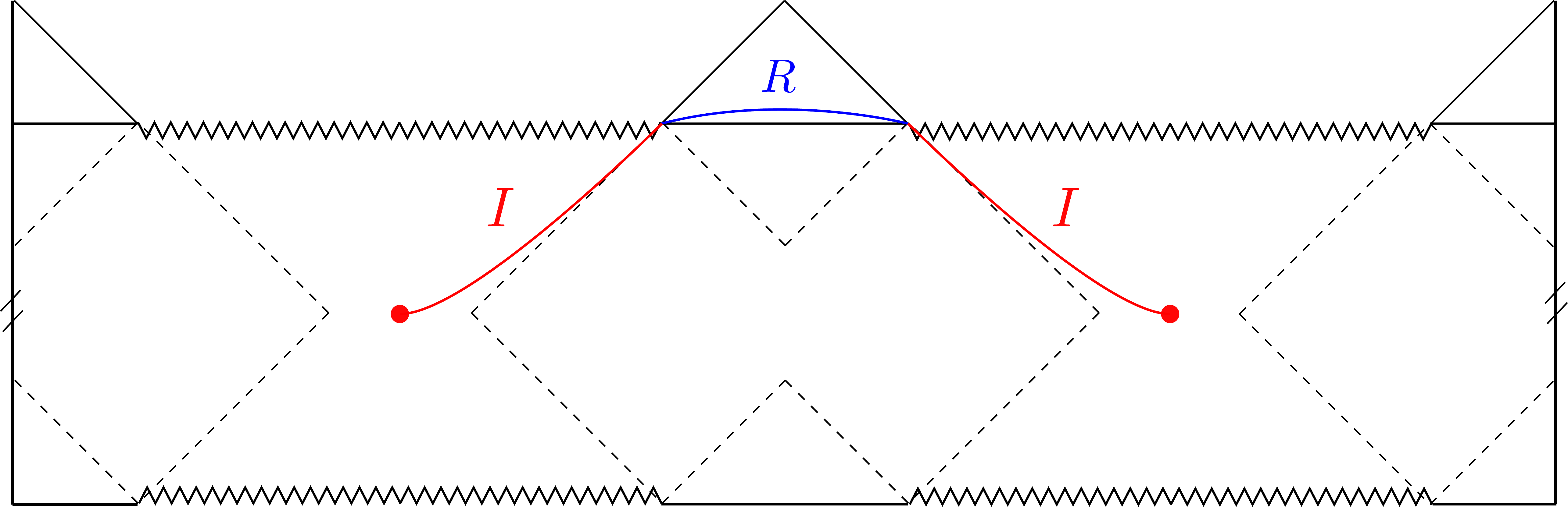}
\caption{The impure saddle for region $R$ with $n=2$.}\label{2hatsqes}
\end{figure}

Having argued that whichever prescription we use to make the connected saddle for $n >1$ dominate the path integral will necessarily preclude the zero-entropy saddle, we are then left with the question of what exactly is the entropy for one of the hats in these connected saddles? In the case of figure \ref{3island}, if the saddle drawn is incorrect, then we are left only with the trivial saddle with vanishing island. Thus the fine-grained entropy of region $R$ is simply the semiclassical matter entropy. But in figure \ref{2hats} we still need to extend the endpoints of region $R$ into the bulk. If they do not run across the entire universe, then where do they end? As it turns out, there exists a nontrivial QES, shown in figure \ref{2hatsqes}.
This can be found explicitly as an extremum of our generalized entropy functional. When $n=2$, it suffices to note that (a) there is a time-reflection $\mathbb{Z}_2$ symmetry around $\s = 0$, (b) the QES for the region $R$ in one hat has to be the same as the QES for the same region in the neighboring hat, by purity of the semiclassical state, and (c) the generalized entropy of a region $R \cup I$ is invariant under $2\pi$ shifts. Fact (a) locates the QES at $\s = 0$, whereas facts (b)-(c) locate the QES at the center of the crunching regions. 

\begin{figure}
\hspace{-10mm}\includegraphics[scale = .2]{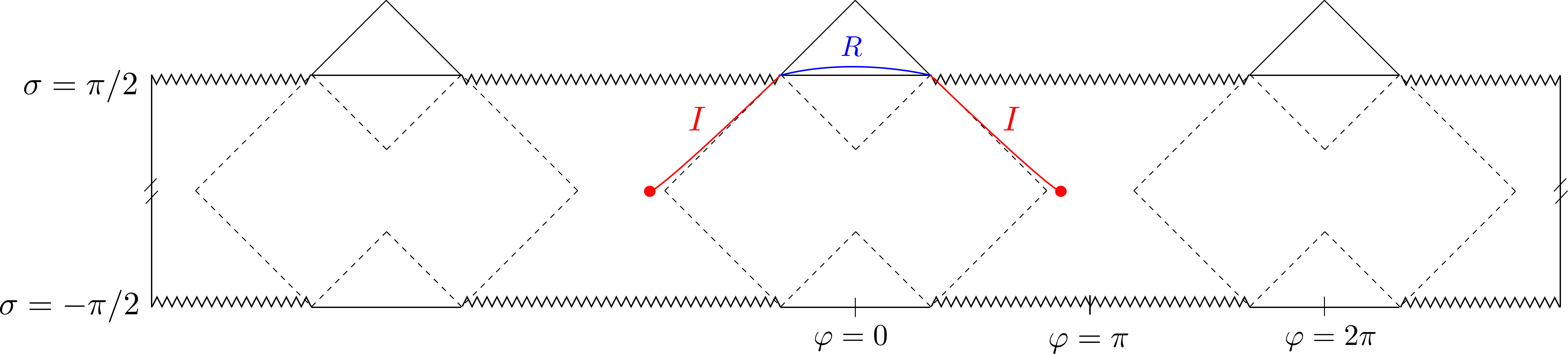}
\caption{The impure saddle for region $R$ with $n=3$.}\label{nontrivial3}
\end{figure}

For $n > 2$ we do not generally have fact (b). But we do know that the center of the crunching regions is classically extremal, so for $\varepsilon=c/\tilde{\phi}_r \ll 1$ the quantum extremal surface will be located nearby. It will still be at $\s = 0$ due to fact (a). By explicit extremization of the generalized entropy functional from a point $\{\s_1, \varphi_1\}$ to $\{\s_2, \varphi_2\}$ in the gravitating part of the spacetime, we find the quantum extremal surface is the pair of points 
\be
\s_1 = \s_2 = 0\,,\qquad \varphi_1 = \pi - \varepsilon, \,\,\varphi_2 = -\pi + \varepsilon\,,\qquad \varepsilon = \f{c \cot \f{\pi}{n}}{6 \tilde{\phi}_r n} + O(c^2/\tilde{\phi}_r^2)\,.
\ee
To the same order in $\varepsilon = c/\tilde{\phi}_r$, the nearby black hole horizons at $\s = 0$ are located at $\varphi =\pm \pi \mp \f{c \pi (1-1/n^2)}{24\tilde{\phi}_r}$. This means that the QES lives beyond the horizon, ensuring that the entanglement wedge of $R$ is larger than its causal past. This is represented for $n=3$ in figure \ref{nontrivial3}. Notice that given the symmetry around $\s = 0$, the growth of the wormhole was necessary to get a nontrivial QES which bounded a spacelike region.

The existence of these saddles is a good sign, because otherwise we would not be able to ascribe an entropy to region $R$. (One option would have been to declare it to be the semiclassical entropy of region $R$, but the fact that its endpoints end at the interface of the gravitating and non-gravitating regions would have been puzzling; in particular we would have to exclude by fiat the gravitational contribution to the entropy here.) We take this as an indication that our setup is self-consistent. 

A somewhat surprising feature of our setup and this result for the entropy is that  it also suggests a fine-grained entropy for a region between the two red QES's, which is \emph{entirely in the gravitationally fluctuating region of the spacetime} (here we are assuming that whatever prescription makes this saddle dominant does not freeze the region $\s < \pi/2$). In other words, we can unitarily deform the region $R \cup I$ in figure \ref{nontrivial3} downward while keeping the red endpoints fixed without changing the fine-grained entropy. This is qualitatively different from the usual scenario in AdS/CFT, where every time slice that includes a Cauchy slice of the entanglement wedge also includes a frozen region. Also, when we normally apply the island rule directly to a region in a gravitating spacetime we find that the region shrinks to zero size and vanishes once we extremize with respect to the region's boundary. In this case we have a ``floating" island which is not anchored to any boundary and does not have an auxiliary region $R$; its endpoints are quantum extremal on their own.  It seems consistent to ascribe such a region a fine-grained entropy equal to its generalized entropy. Of course, even if this is possible it is related to the fact that gravity was frozen somewhere else in the spacetime. 

Note that if we froze spacetime to the past of $\s = \pi/2$ as in section \ref{sec:frozen}, then unitarily deforming $R \cup I$ downward from the hat would not lead to any surprises, since any unitary deformation of the region would include a frozen region.

\section{Discussion}\label{sec:discussion}
The observer-dependence of cosmic horizons makes the concept of encoding the region beyond the horizon more subtle than in the context of black holes. In particular, two observers outside of a black hole will agree on the black hole event horizon, and only one of them will be able to encode the interior. However, two observers in different places in the universe will have different cosmic horizons: if they each encode the region beyond their horizon, then those regions can overlap. If this were realized it would lead to inconsistencies with quantum mechanics. 

The resolution to this problem is quite simple in our setup. It depends on  carefully defining the microscopic description. Our microscopic description is given by a CFT on the ``frozen" regions (i.e. the regions where the quantum effects of gravity are ignored) with a boundary condition at an initial time. The initial boundary condition is prepared by a gravitational path integral, and in this paper we have looked for various saddles which dominate this path integral. To reach a potential paradox, we considered freezing two disconnected regions, corresponding to two different observers. Now, if we do nothing to entangle these two regions in the microscopic description, then the dominant saddle which fills them in will correspond to two distinct universes, one for each frozen region. In such a situation we will not have overlapping entanglement wedges, and therefore no paradox (each observer will encode the region beyond their horizon but within their connected piece of the universe). However, if we entangle the two frozen regions in the microscopic description, then we can have the leading saddle be a single universe hosting both frozen regions. But entangling the two frozen regions has to be a process which references both, so in a computation of the purity, the replica wormhole which would give a pure state for either frozen region is disallowed. It is disallowed since it would require swapping one of the two frozen hats with its replica, but this would violate the procedure that entangled the two regions, e.g. by the procedure in \ref{sec:frozen}. The entanglement wedge of either frozen region does not run across the entire universe; we instead have complementary entanglement wedges for each frozen region, as expected.

\begin{figure}
\centering
\includegraphics[scale = .2]{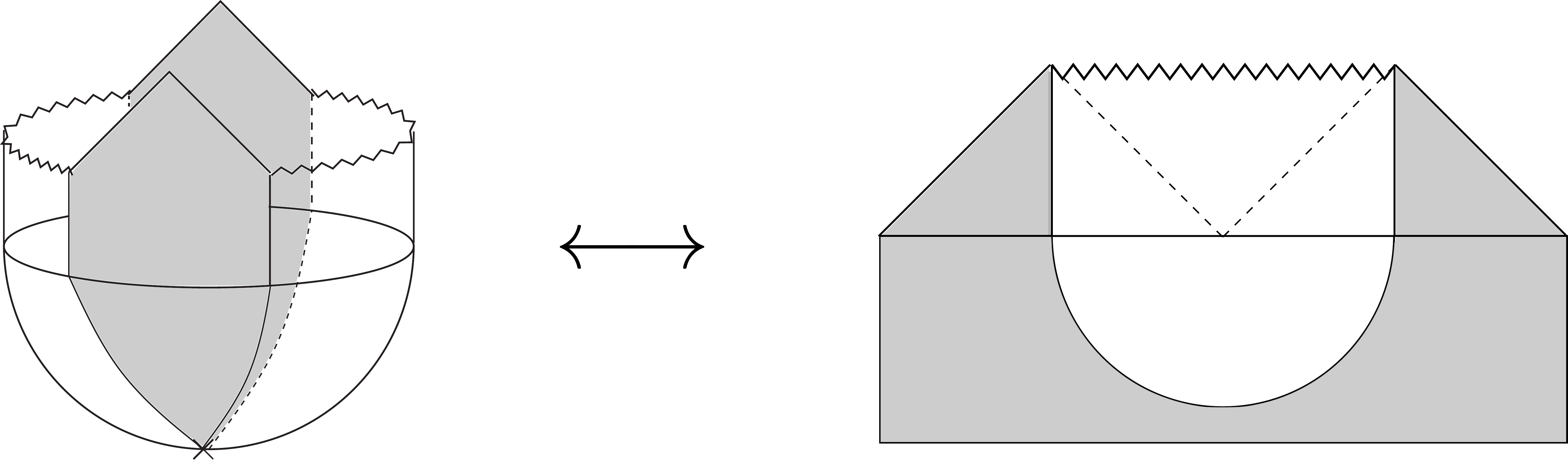}\vspace{2mm}
\caption{On the left we have reproduced figure \ref{fig:twohat3dfrozen}, corresponding to the Euclidean preparation of a $4\pi$ universe with frozen regions denoted in gray. On the right we have the Euclidean preparation of the thermofield double in AdS coupled to frozen flat-space ``wing" regions. The solid black line in the middle is the time-reflection symmetric point where the Euclidean manifold is pasted to the Lorentzian manifold. These two situations are analogous.}\label{tfd}
\end{figure}

\begin{figure}
\centering
\includegraphics[scale = .18]{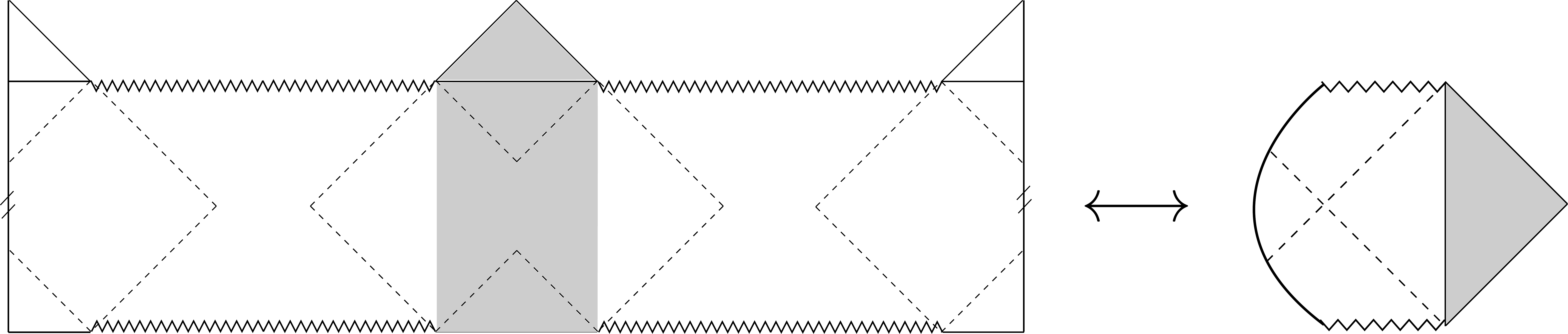}\vspace{2mm}
\caption{On the right we have a pure-state black hole with an end-of-the-world brane behind the horizon. This means that the single wing system (plus boundary) encodes the entire bulk, and has fine-grained entropy $S = 0$. This corresponds to unfreezing one of the two frozen regions in the TFD in figure \ref{tfd}. It is analogous to a single frozen hat region in dS, shown on the left. In this case the theory in the hat has fine-grained entropy $S = 0$ and encodes the entire universe on its own.}\label{eow}
\end{figure}

It is important in the situations we have analyzed that the frozen regions define the microscopic description, and therefore the entropy. This is true even if the frozen regions are somewhere else in the universe. A simple analogy is the thermofield double (TFD) in AdS, coupled to non-gravitating flat-space wings. This is prepared by the Euclidean path integral on the right-hand-side of figure \ref{tfd}. Notice the left and right wings, which would have been disconnected in the Lorentzian manifold, are connected by a frozen region in the Euclidean manifold. This is just as in figure \ref{fig:twohat3dfrozen} with our frozen flat-space hats, reproduced on the left-hand-side of figure \ref{tfd} (the two shield-like regions meet in the vicinity of the conical singularity in the Euclidean past). So in the computation of the entropy of a hat or wing using the island rule, the region cannot be extended to the entire universe since it will run into another frozen region. This means $S \neq 0$. In a replica computation, like in section \ref{sec:frozen}, the connected frozen regions must remain connected, disallowing the purity saddle with $S = 0$ which swaps hats/wings. In the TFD the entropy of either the left or right system is given by the black hole entropy. We saw a similar result for the entropy of a single hat in section \ref{hatentropy}. However, the two wings, or two hats, can be disentangled, by unfreezing one of the two. In the TFD this can be done by inserting an end-of-the-world brane behind the horizon. Then the full microscopic description is just given by one side, and its entropy now vanishes, see the right-hand-side of figure \ref{eow}. This corresponds to ``unfreezing" the left region, in which case there is no problem with the right region encoding the interior (and what used to be the left exterior) on its own. This would be the same as if we only froze one of the two hat regions, shown on the left-hand-side of figure \ref{eow}: the entropy of the theory in the hat now vanishes, and it encodes the entire universe on its own. (The universe drawn in figure \ref{eow} is size $4\pi$ because we imagine the frozen region includes the conical singularity of particular opening angle in the Euclidean past, which fixes the size of the universe. If it were not included then the frozen hat would like in a $2\pi$ universe.)

A complicating feature of our analysis was the conical singularity in Euclidean signature that prohibits a conventional preparation of the Hartle-Hawking state. One option to deal with this, although there is not an obvious candidate, is to find a bra-ket wormhole in global-like coordinates which avoids the universe capping off in the Euclidean section. This would avoid the conical singularity.

\subsection{Comments on quantum cosmology}
The fact that our observers tend to split up into disconnected universes depends crucially on using the Hartle-Hawking wavefunction. Proposals like Vilenkin's tunneling wavefunction \cite{Vilenkin:1983xq} seem to give the inverse answer for the amplitude in these simple setups \cite{Vilenkin:1984wp, Feldbrugge:2017kzv}. In such a situation we would find a preference for our observers to be in the same connected piece of the universe. It would be interesting to analyze whether there remains a paradox in this case. 

Another interesting aspect of our analysis is that we see a semiclassical avatar of the picture advocated in \cite{Hartle:2010dq, Hartle:2016tpo, Aguilar-Gutierrez:2021bns}, where one can think of distinct observers as living in their own universe due to coarse-graining beyond their horizons.

\subsection{Encoding the $n>1$ universe in a CFT$_2$?}
\begin{figure}
\begin{subfigure}{.4\linewidth}
\centering
\includegraphics[width=.5\linewidth]{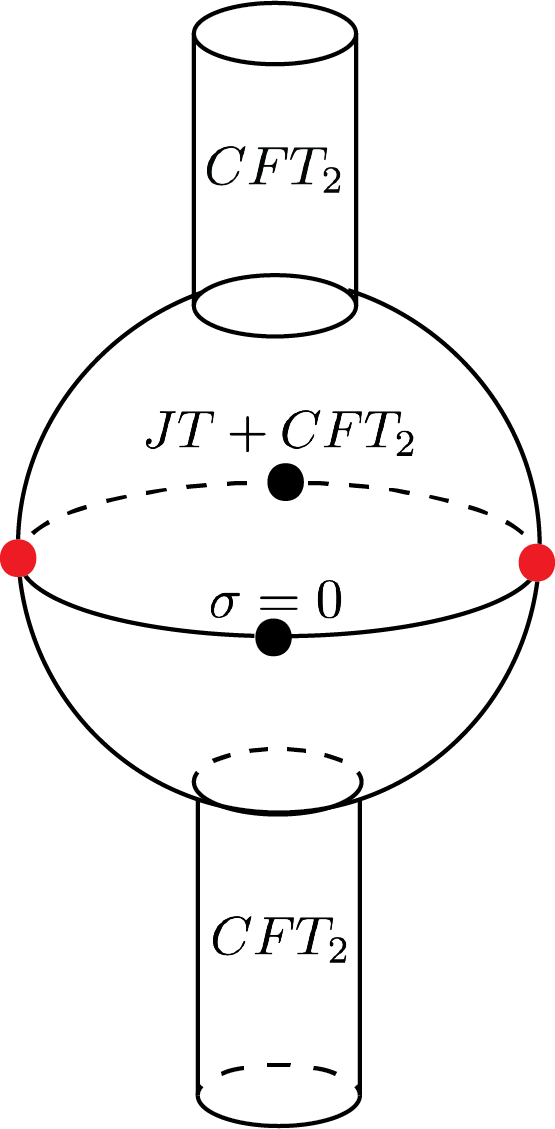}
\caption{}
\label{fig:CFTvac}
\end{subfigure}
\begin{subfigure}{.4\linewidth}
\centering
\includegraphics[width=.8\linewidth]{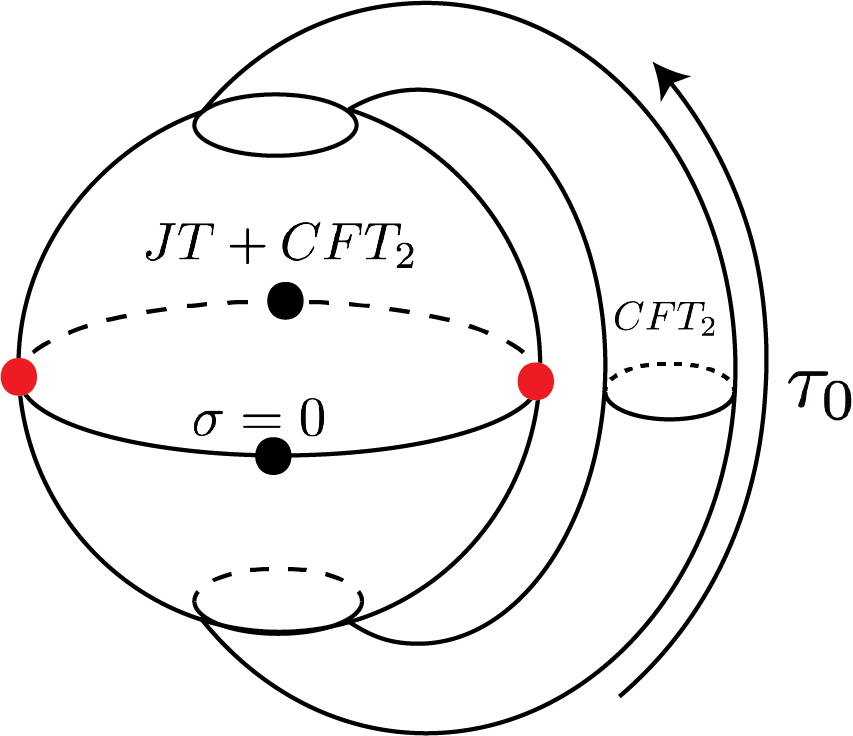}
\caption{}
\label{fig:CFTthermal}
\end{subfigure}
\setlength{\belowcaptionskip}{-10pt}
\caption{In the figure on the left, we imagine attaching two semi-infinite cylinders to holes at the north and south poles of the sphere. Here we have exaggerated the size of the holes relative to the size of the sphere. This prepares the CFT matter state in the vacuum on the circle, as was considered in section \ref{sec:setup}. On the right, we imagine making the CFT state thermal. This will raise the energy in the spacetime and make the wormholes longer upon analytic continuation. The red dots denote points of maximal dilaton and the black dots denote points of minimal dilaton along the $\sigma =0$ time slice.}
\label{fig:CFTencode}
\end{figure}

A curious fact about our solution described in section \ref{sec:setup} is that after accounting for the back-reaction from the quantum matter, our dilaton solution actually blows up to $+\infty$ as one approaches the conical singularity. To see this, recall that the backreacted dilaton takes the form in global dS$_2$ coordinates
\begin{align}
   \phi(\sigma, \varphi) = \tilde{\phi}_r \frac{\cos \varphi}{\cos \sigma} - \frac{c}{12} \left(1- \frac{1}{n^2}\right)\sigma \tan \sigma.
\end{align}
Continuing $\sigma \to is$ and taking $s \to \pm \infty$, we see that the dilaton actually blows up linearly in $s$ to $+\infty$. This means that gravitational effects are becoming weak near the north and south poles of the sphere. It is natural to introduce another boundary at these locations to cut off the spacetime. 
In particular, we could imagine cutting out small holes around the conical singularities and gluing in two semi-infinite cylinders, each ending on one of the holes. We can imagine the bulk matter CFT continues to propagate along these cylinders. This is illustrated in figure \ref{fig:CFTvac}. 

\begin{figure}
\centering
\includegraphics[width=.5\textwidth]{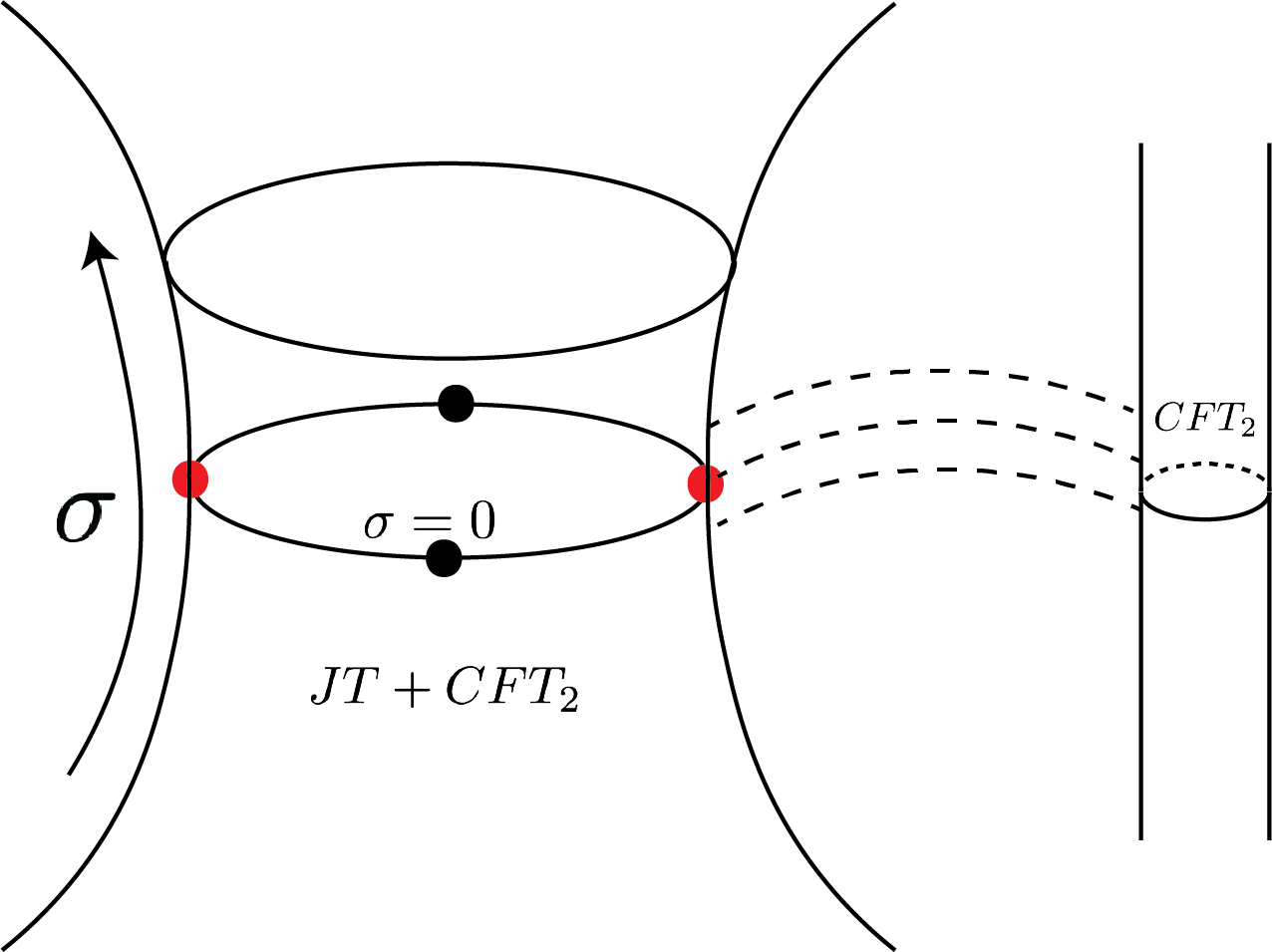}
\caption{Here we illustrate the Lorentzian interpretations of the Euclidean saddles shown in figure \ref{fig:CFTencode}. The idea is that for finite $\tau_0$ these saddles describe a situation where a closed dS$_2$ universe with multiple inflating and crunching regions is encoded in a CFT$_{2}$ on a spatial circle. The dashed lines signify the thermal entanglement between the closed universe and the CFT.}\label{fig:CFTLorentzian}
\end{figure}

We could further imagine connecting these two cylinders together to form a portion of a torus, with some length $\tau_0$. Doing so modifies the stress energy in the bulk portion of the spacetime, putting the quantum fields in a thermal state with temperature set by $\tau_0$. This is also illustrated in figure \ref{fig:CFTthermal}. The resulting bulk saddle will still have a moment of time-reflection symmetry at $s=i\sigma =0$. We can then analytically continue the saddle from this moment of symmetry and the picture we get is that of multiple inflating patches connected via wormholes similar to the saddles discussed in this work, where the full universe is entangled with a 2d CFT on a circle. This is illustrated in figure \ref{fig:CFTLorentzian}. By the island rule, the microscopic entropy of the CFT is zero, since the island just includes the full closed universe. In other words, the CFT will encode certain observables in the inflating patches. This part is not surprising, and has been explored in previous work \cite{Almheiri:2019tt, Balasubramanian:2020ue}, but a construction like the one above would allow us to probe subsystem encoding, i.e. to see if subregions of the CFT$_2$ encode subregions of the de Sitter universe.\footnote{What one would like is an island region that includes part of the inflating patch of the spacetime. But insofar as the endpoints of this region are quantum extremal cousins of the classically extremal cosmic horizon, one will run into issues with entanglement wedge nesting \cite{Shaghoulian:2022aa}. The basic issue is that the bifurcate cosmic horizon is a ``minimax" surface (minimum in time, maximum in space) as opposed to a maximin surface like the bifurcate black hole horizon. Here there would be both cosmic and black hole horizons in the closed universe and so there is hope of finding a maximin QES in the closed universe, presumably close to the black dots in figure \ref{fig:CFTLorentzian}.} This encoding is similar in spirit to what happens for the encoding of the black hole interior in the radiation after a black hole has fully evaporated and also to scenarios recently discussed in \cite{Chen:2020tes, Cooper:2018aa, Raamsdonk:2021aa, Antonini:2022aa}. The difference for us is that now we have the possibility of encoding asymptotically inflating regions in the dual CFT. This is reminiscent of proposals for ``making a universe in the lab'' by creating inflating regions behind the black hole horizon in AdS/CFT \cite{Freivogel_2006}.\footnote{The authors of \cite{Freivogel_2006} argued against the boundary CFT describing the inflating region. Their claim was that the boundary CFT had to be in a mixed state, obtained by tracing out the degrees of freedom which describe the inflating region. With only causal wedge reconstruction, this is a reasonable conclusion. But given the modern understanding of entanglement wedge reconstruction -- and in particular the encoding of ``bag-of-gold" type geometries (like an inflating region behind a black hole horizon) -- there is no obstruction to such a configuration being described by a pure state in the boundary CFT.} It would be interesting to understand this encoding in more detail and we hope to do so in the future. 

\bigskip\bigskip\bigskip

\noindent \textbf{Acknowledgments} We would like to thank Raphael Bousso, Yiming Chen, Daniel Harlow, Tom Hartman, Juan Maldacena, Don Marolf and Henry Maxfield for useful conversations. ES is supported by the Simons Foundation It from Qubit collaboration (385592) and the DOE QuantISED grant DESC0020360. AL acknowledges support from NSF grant PHY-1911298 and Carl P. Feinberg. This research was supported in part by the National Science Foundation under Grant No. NSF PHY-1748958. This work was performed in part at the Aspen Center for Physics, which is supported by National Science Foundation grant PHY-1607611.

\appendix
\section{Casimir energy on $S^1 \times S^2$}\label{app:casimir}
Here we adapt an argument from \cite{Belin_2017} to derive some constraints on the vacuum energy of a conformal field theory on $S^1_L \times S^2$ as a function of the size $L$. The radius of the $S^2$ will be denoted by $R$ but scale-invariant quantities will depend on the ratios $R/L$. The key tool will be SL$(2,\mathbb{Z})$ invariance of the thermal partition function on $S^1_L \times S^2$:
\be
Z(\beta)_{S^1_L \times S^2} = Z(L)_{S^1_\beta \times S^2}
\ee
Denote $V =  \text{Vol}(S^1_\b \times S^1_L \times S^2)$. To extract the vacuum energy density in the original channel we go to low temperatures:
\be\label{channel1}
\lim_{\beta \rightarrow \infty} V^{-1} \log Z(\beta)_{S^1_L \times S^2} \approx -\f{ \b E_{\text{vac}}}{V} = -R^{-4} g(L/R)
\ee
Defining $\D = E-E_{\text{vac}}$ we now consider the dual channel at finite $L$:
\begin{align}
\lim_{\beta \rightarrow \infty} \,\,V^{-1}\log Z(L)_{S^1_\b \times S^2} &= \lim_{\beta \rightarrow \infty} \,\,V^{-1}\log \left(\sum \exp\left(- L E\right)\right)\\
& = \lim_{\beta \rightarrow \infty} \,\,V^{-1}\log \left(\exp(-L E_{\text{vac}})  \sum \exp\left(- L \D\right)\right)\\
& = \lim_{\beta \rightarrow \infty}\left(-R^{-4} g(\b/R) + V^{-1}\log \left(\sum \exp(-L\D)\right)\right)\label{channel2}
\end{align}
Equating \eqref{channel1} with \eqref{channel2} and taking an $L$ derivative gives
\be
-\partial_L\, R^{-4} g(L/R) = \partial_L \,\lim_{\b \rightarrow \infty} \,\left(\f{1}{4\pi R^2 \b L} \log\left(\sum \exp(-L \D)\right)\right)
\ee
 This derivative is negative by unitarity $\D>0$, so we find
\be
\partial_L \f{E_{\text{vac},S^1_L \times S^2}}{4\pi R^2 L} = \partial_L R^{-4} g(L/R) \geq 0 \,,
\ee
i.e. increasing the size of the spatial $S^1$ compared to the $S^2$ increases the energy density. This is all we need in the main text, but we can say more. By extensivity of the thermal entropy in the dual channel we have
\be
\lim_{L \rightarrow 0} \,\,\lim_{\b \rightarrow \infty}\,\,V^{-1} \log Z(\beta)_{S^1_L \times S^2}= \lim_{L \rightarrow 0}\,\,\lim_{\b \rightarrow \infty}\,\,V^{-1} \log Z(L)_{S^1_\b \times S^2}\approx  \f{c}{L^4}\,.
\ee
This means $g(L/R \rightarrow 0) < 0$. We can also provide a sign constraint for the second derivative. We will drop some irrelevant prefactors and the large-$\b$ limit. We have
\be
\partial_L^2 \left(\f{1}{L} \log\left(\sum e^{-L \D}\right)\right) = \f{2}{L^3}\log\left(\sum e^{-L \D}\right) +\f{1}{L^2} \f{\sum \D e^{-L \D}}{\sum e^{-L \D}}+\f{1}{L}\partial_L^2 \log\left(\sum e^{-L \D}\right)
\ee
The first two terms are manifestly positive and the last term can be rewritten as 
\begin{align}
\f{1}{L} \partial_L \left(\f{- \sum_i \D_i e^{-L \D_i}}{\sum_j e^{-L \D_j}}\right) &= \f{\left(\sum_i \D_i^2 e^{-L \D_i}\right)\left(\sum_j e^{-L \D_j}\right) - \left(\sum_i \D_i e^{- L \D_i}\right)^2}{L \left(\sum_i e^{- L \D_i}\right)^2}\\
&=\f{\sum_{ij} e^{-L (\D_i + \D_j)}(\D_i - \D_j)^2}{2L\left(\sum_i e^{- L \D_i}\right)^2} \geq 0\,.
\end{align}
This tells us that $\partial_L^2 g(L/R) \leq 0$. So altogether we have
\be
\boxed{ g(L/R \rightarrow 0) = - c\f{R^4}{L^4}\,,\qquad \partial_L g(L/R) \geq 0\,,\qquad \partial_L^2 g(L/R) \leq 0}
\ee
This means that $g(L/R \rightarrow \infty)$ must approach a finite constant, as long as the second derivative inequality is strict for large $L/R$. This is reasonable, since the only parameter the vacuum energy density can depend on in this limit is $R$, and by dimensional analysis the expression should be $k/R^4$ for some constant $k = g(L/R \rightarrow \infty)$. 

\section{Coordinates on Milne wedge}\label{app:coords}
In this appendix we exhibit the coordinates on the Milne-like wedge which covers the causal past of the portion of $\mathcal{I}^+$ that is inflating. We are working with a $2\pi n$-sized universe with $n>1$. Since the local metric does not depend on $n$ we will use the isometries of dS$_2$ to shrink the Milne wedge of the $n=1$ case. The isometries of dS$_2$ are best understood through the embedding space,
\begin{align}
ds^2 = -dX_0^2 + dX_1^2 + dX_2^2\,,\qquad -X_0^2 + X_1^2 + X_2^2 = 1\\
X_0 = \tan \s\,,\quad X_1 = \f{\sin \varphi}{\cos \s}\,,\quad X_2 = \f{\cos \varphi}{\cos \s}\label{globemb}\,.
\end{align}
The isometries are given by the isometries of the embedding space $\mathbb{R}^{1,2}$ which preserve the dS$_2$ hyperboloid. We therefore have the two boosts and rotation generators
\be
K_i = X_0 \p_i + X_i \p_0\,,\qquad J_{12} = X_1 \p_2 - X_2 \p_1
\ee
Shrinking the Milne wedge will require fiddling with the time coordinate, so we will require one of the two boosts, which in global coordinates are 
\be
K_1 = \sin \s \cos\varphi\, \p_\varphi + \cos \s \sin \varphi\, \p_\s\,,\qquad K_2 = \cos \s \cos \varphi \,\p_\s - \sin \s \sin\varphi\, \p_\varphi\,.
\ee
We see that $K_2$ shifts $\s = \varphi = 0$ upward, which is what we need to shrink the Milne wedge, while $K_1$ keeps this point fixed. In fact $K_2$ shrinks the entire Milne wedge as needed. We therefore apply it as $\L = \exp(\a_n K_2)$. Since it is a Lorentz boost it will act on the embedding coordinates as 
\be
\begin{pmatrix} X_0 \\ X_2\end{pmatrix} \longrightarrow \f{1}{\sqrt{1-\a_n^2}}\begin{pmatrix} X_0 - \a_n X_2\\X_2 - \a_n X_0\end{pmatrix} 
\ee
We set these equal to new Milne coordinates $\tilde{X_0} = -\f{\cosh \tilde{\chi}}{\sinh \tilde{\eta}}$, $\tilde{X}_1 = -\f{\sinh \tilde{\chi}}{\sinh \tilde{\eta}}$, $\tilde{X}_2 = -\coth \tilde{\eta}$ and solve for $\tilde{\eta}$, $\tilde{\chi}$ to get 
\begin{align}
    &\tanh(\tc) = \sqrt{1-\alpha_n^2}\, \frac{\sin \varphi}{\sin \sigma - \alpha_n \cos \varphi} \nonumber \\
    & \tanh(\te) =\sqrt{1-\alpha_n^2} \,\frac{\cos \sigma}{\alpha_n \sin \sigma -\cos \varphi}.
\end{align}

\section{Thermal coordinates on Milne wedge}\label{app:coordstherm}
We begin by writing down the coordinate transformation between the Poincar\'e patch and the global cylinder.  The embedding coordinates for the Poincar\'e patch are given by
\be
X_0 = \f{\eta^2 - x^2 - 1}{2\eta}\,,\qquad X_1 = \f{x}{\eta}\,,\qquad X_2 = \f{\eta^2 - x^2 + 1}{2\eta}\,.
\ee
Equating this with \eqref{globemb} gives
\be
\eta = \f{\sin \s}{\cos\varphi + \cos \s}\,,\qquad x = \f{\sin \varphi}{\cos\varphi + \cos\s}\,.
\ee
In our case, we want a cylinder that is $n$ times bigger, and for convenience we want $\s = 0$ to map to $\eta = 0$, so we will shift/rescale $\{\s, \varphi\}\rightarrow \{(\s-\pi/2)/n,\varphi/n\}$.  The region $\s = \pi/2, \varphi \in (-\varphi_*, \varphi_*)$ maps to  $\eta = 0, x \in (-x_*, x_*)$ with $x_* = \sin (\varphi_*/n)/((\cos\varphi_*/n)+1)$. Performing an SL$(2)$ transformation to map this to the half-line $x\in (0, \infty)$ gives 
\be
t_{\text{half}} \pm x_{\text{half}}=\mp  \f{\eta \pm x + x_*}{\eta \pm x - x_*}
\ee
Now going to thermal Rindler-like coordinates gives
\be
t_{\text{half}} \pm x_{\text{half}} = \pm e^{t_{\text{th}} \pm x_{\text{th}}}
\ee
Solving for $t_{\text{th}}$, $x_{\text{th}}$ gives
\begin{align}
    &\tanh x_{\text{th}} = \sqrt{1-\beta_n^2} \frac{\sin \frac{\varphi}{n}}{\cos \frac{\sigma - \pi/2}{n} - \beta_n \cos \frac{\varphi}{n}}\nonumber \\
    &\tanh t_{\text{th}} = -\sqrt{1-\beta_n^2} \frac{\sin \frac{\sigma - \pi/2}{n}}{\beta_n \cos \frac{\sigma - \pi/2}{n} - \cos \frac{\varphi}{n}} 
\end{align}
where $\beta_n = \cos \frac{\varphi_*}{n}$, recovering \eqref{thermcoords}. 

\footnotesize
\bibliographystyle{ourbst}
\bibliography{dS2multibib}

\providecommand{\href}[2]{#2}\begingroup\raggedright\begin{thebibliography}{10}

\bibitem{Gibbons:1976ue}
G.~W. Gibbons and S.~W. Hawking, ``{Action Integrals and Partition Functions in
  Quantum Gravity},'' \href{http://dx.doi.org/10.1103/PhysRevD.15.2752}{{\em
  Phys. Rev. D} {\bfseries 15} (1977) 2752--2756}.

\bibitem{Freivogel:2006xu}
B.~Freivogel, Y.~Sekino, L.~Susskind, and C.-P. Yeh, ``{A Holographic framework
  for eternal inflation},''
  \href{http://dx.doi.org/10.1103/PhysRevD.74.086003}{{\em Phys. Rev. D}
  {\bfseries 74} (2006) 086003},
  \href{http://arxiv.org/abs/hep-th/0606204}{{\ttfamily arXiv:hep-th/0606204}}.

\bibitem{Chen:2020tes}
Y.~Chen, V.~Gorbenko, and J.~Maldacena, ``{Bra-ket wormholes in gravitationally
  prepared states},'' \href{http://dx.doi.org/10.1007/JHEP02(2021)009}{{\em
  JHEP} {\bfseries 02} (2021) 009},
  \href{http://arxiv.org/abs/2007.16091}{{\ttfamily arXiv:2007.16091
  [hep-th]}}.

\bibitem{Hartman:2020khs}
T.~Hartman, Y.~Jiang, and E.~Shaghoulian, ``{Islands in cosmology},''
  \href{http://dx.doi.org/10.1007/JHEP11(2020)111}{{\em JHEP} {\bfseries 11}
  (2020) 111}, \href{http://arxiv.org/abs/2008.01022}{{\ttfamily
  arXiv:2008.01022 [hep-th]}}.

\bibitem{Maldacena:2002vr}
J.~M. Maldacena, ``{Non-Gaussian features of primordial fluctuations in single
  field inflationary models},''
  \href{http://dx.doi.org/10.1088/1126-6708/2003/05/013}{{\em JHEP} {\bfseries
  05} (2003) 013}, \href{http://arxiv.org/abs/astro-ph/0210603}{{\ttfamily
  arXiv:astro-ph/0210603}}.

\bibitem{Strominger:2001pn}
A.~Strominger, ``{The dS / CFT correspondence},''
  \href{http://dx.doi.org/10.1088/1126-6708/2001/10/034}{{\em JHEP} {\bfseries
  10} (2001) 034}, \href{http://arxiv.org/abs/hep-th/0106113}{{\ttfamily
  arXiv:hep-th/0106113}}.

\bibitem{Bousso:1999dw}
R.~Bousso, ``{The Holographic principle for general backgrounds},''
  \href{http://dx.doi.org/10.1088/0264-9381/17/5/309}{{\em Class. Quant. Grav.}
  {\bfseries 17} (2000) 997--1005},
  \href{http://arxiv.org/abs/hep-th/9911002}{{\ttfamily arXiv:hep-th/9911002}}.

\bibitem{Banks:2000fe}
T.~Banks, ``{Cosmological breaking of supersymmetry?},''
  \href{http://dx.doi.org/10.1142/S0217751X01003998}{{\em Int. J. Mod. Phys. A}
  {\bfseries 16} (2001) 910--921},
  \href{http://arxiv.org/abs/hep-th/0007146}{{\ttfamily arXiv:hep-th/0007146}}.

\bibitem{Fischler}
W.~Fischler, ``{Taking de Sitter seriously. Talk given at Role of scaling laws
  in physics and biology (Celebrating the 60th birthday of Geoffrey West)},''
  {\em Santa Fe, Dec} (2000) .

\bibitem{Bousso:2000nf}
R.~Bousso, ``{Positive vacuum energy and the N bound},''
  \href{http://dx.doi.org/10.1088/1126-6708/2000/11/038}{{\em JHEP} {\bfseries
  11} (2000) 038}, \href{http://arxiv.org/abs/hep-th/0010252}{{\ttfamily
  arXiv:hep-th/0010252}}.

\bibitem{Banks:2001yp}
T.~Banks and W.~Fischler, ``{M theory observables for cosmological
  space-times},'' \href{http://arxiv.org/abs/hep-th/0102077}{{\ttfamily
  arXiv:hep-th/0102077}}.

\bibitem{Parikh:2002py}
M.~K. Parikh, I.~Savonije, and E.~P. Verlinde, ``{Elliptic de Sitter space:
  dS/Z(2)},'' \href{http://dx.doi.org/10.1103/PhysRevD.67.064005}{{\em Phys.
  Rev. D} {\bfseries 67} (2003) 064005},
  \href{http://arxiv.org/abs/hep-th/0209120}{{\ttfamily arXiv:hep-th/0209120}}.

\bibitem{Dyson:2002nt}
L.~Dyson, J.~Lindesay, and L.~Susskind, ``{Is there really a de Sitter/CFT
  duality?},'' \href{http://dx.doi.org/10.1088/1126-6708/2002/08/045}{{\em
  JHEP} {\bfseries 08} (2002) 045},
  \href{http://arxiv.org/abs/hep-th/0202163}{{\ttfamily arXiv:hep-th/0202163}}.

\bibitem{Banks:2002wr}
T.~Banks, W.~Fischler, and S.~Paban, ``{Recurrent nightmares? Measurement
  theory in de Sitter space},''
  \href{http://dx.doi.org/10.1088/1126-6708/2002/12/062}{{\em JHEP} {\bfseries
  12} (2002) 062}, \href{http://arxiv.org/abs/hep-th/0210160}{{\ttfamily
  arXiv:hep-th/0210160}}.

\bibitem{Banks:2003pt}
T.~Banks and W.~Fischler, ``{An Upper bound on the number of e-foldings},''
  \href{http://arxiv.org/abs/astro-ph/0307459}{{\ttfamily
  arXiv:astro-ph/0307459}}.

\bibitem{Banks:2006rx}
T.~Banks, B.~Fiol, and A.~Morisse, ``{Towards a quantum theory of de Sitter
  space},'' \href{http://dx.doi.org/10.1088/1126-6708/2006/12/004}{{\em JHEP}
  {\bfseries 12} (2006) 004},
  \href{http://arxiv.org/abs/hep-th/0609062}{{\ttfamily arXiv:hep-th/0609062}}.

\bibitem{Banks:2015iya}
T.~Banks and W.~Fischler, {\em {Holographic Inflation Revised}},
  \href{http://dx.doi.org/10.1017/9781316535783.013}{pp.~241--262}.
\newblock 2017.
\newblock \href{http://arxiv.org/abs/1501.01686}{{\ttfamily arXiv:1501.01686
  [hep-th]}}.

\bibitem{Banks:2018ypk}
T.~Banks and W.~Fischler, ``{The holographic spacetime model of cosmology},''
  \href{http://dx.doi.org/10.1142/S0218271818460057}{{\em Int. J. Mod. Phys. D}
  {\bfseries 27} no.~14, (2018) 1846005},
  \href{http://arxiv.org/abs/1806.01749}{{\ttfamily arXiv:1806.01749
  [hep-th]}}.

\bibitem{Banks:2020zcr}
T.~Banks and W.~Fischler, ``{Holographic Space-time, Newton`s Law, and the
  Dynamics of Horizons},'' \href{http://arxiv.org/abs/2003.03637}{{\ttfamily
  arXiv:2003.03637 [hep-th]}}.

\bibitem{Susskind:2021yvs}
L.~Susskind, ``{Three Impossible Theories},''
  \href{http://arxiv.org/abs/2107.11688}{{\ttfamily arXiv:2107.11688
  [hep-th]}}.

\bibitem{Shaghoulian:2021cef}
E.~Shaghoulian, ``{The central dogma and cosmological horizons},''
  \href{http://dx.doi.org/10.1007/JHEP01(2022)132}{{\em JHEP} {\bfseries 01}
  (2022) 132}, \href{http://arxiv.org/abs/2110.13210}{{\ttfamily
  arXiv:2110.13210 [hep-th]}}.

\bibitem{Anninos:2011jp}
D.~Anninos, G.~S. Ng, and A.~Strominger, ``{Future Boundary Conditions in De
  Sitter Space},'' \href{http://dx.doi.org/10.1007/JHEP02(2012)032}{{\em JHEP}
  {\bfseries 02} (2012) 032}, \href{http://arxiv.org/abs/1106.1175}{{\ttfamily
  arXiv:1106.1175 [hep-th]}}.

\bibitem{Aguilar-Gutierrez:2021bns}
S.~E. Aguilar-Gutierrez, A.~Chatwin-Davies, T.~Hertog, N.~Pinzani-Fokeeva, and
  B.~Robinson, ``{Islands in multiverse models},''
  \href{http://dx.doi.org/10.1007/JHEP11(2021)212}{{\em JHEP} {\bfseries 11}
  (2021) 212}, \href{http://arxiv.org/abs/2108.01278}{{\ttfamily
  arXiv:2108.01278 [hep-th]}}.

\bibitem{Kay:1988mu}
B.~S. Kay and R.~M. Wald, ``{Theorems on the Uniqueness and Thermal Properties
  of Stationary, Nonsingular, Quasifree States on Space-Times with a Bifurcate
  Killing Horizon},''
  \href{http://dx.doi.org/10.1016/0370-1573(91)90015-E}{{\em Phys. Rept.}
  {\bfseries 207} (1991) 49--136}.

\bibitem{Harlow_2017}
D.~Harlow, ``The ryu{\textendash}takayanagi formula from quantum error
  correction,'' \href{https://doi.org/10.1007%2Fs00220-017-2904-z}{{\em
  Communications in Mathematical Physics} {\bfseries 354} no.~3, (May, 2017)
  865--912}.

\bibitem{Almheiri_2021}
A.~Almheiri, T.~Hartman, J.~Maldacena, E.~Shaghoulian, and A.~Tajdini, ``The
  entropy of hawking radiation,''
  \href{https://doi.org/10.1103%2Frevmodphys.93.035002}{{\em Reviews of Modern
  Physics} {\bfseries 93} no.~3, (Jul, 2021) }.

\bibitem{Akers:2018aa}
C.~Akers and P.~Rath, ``Holographic renyi entropy from quantum error
  correction,'' \href{http://arxiv.org/abs/1811.05171}{{\ttfamily 1811.05171}}.
  \url{https://arxiv.org/pdf/1811.05171.pdf}.

\bibitem{Dong:2018aa}
X.~Dong, D.~Harlow, and D.~Marolf, ``Flat entanglement spectra in fixed-area
  states of quantum gravity,''
  \href{http://arxiv.org/abs/1811.05382}{{\ttfamily 1811.05382}}.
  \url{https://arxiv.org/pdf/1811.05382.pdf}.

\bibitem{Vilenkin:1983xq}
A.~Vilenkin, ``{The Birth of Inflationary Universes},''
  \href{http://dx.doi.org/10.1103/PhysRevD.27.2848}{{\em Phys. Rev. D}
  {\bfseries 27} (1983) 2848}.

\bibitem{Vilenkin:1984wp}
A.~Vilenkin, ``{Quantum Creation of Universes},''
  \href{http://dx.doi.org/10.1103/PhysRevD.30.509}{{\em Phys. Rev. D}
  {\bfseries 30} (1984) 509--511}.

\bibitem{Feldbrugge:2017kzv}
J.~Feldbrugge, J.-L. Lehners, and N.~Turok, ``{Lorentzian Quantum Cosmology},''
  \href{http://dx.doi.org/10.1103/PhysRevD.95.103508}{{\em Phys. Rev. D}
  {\bfseries 95} no.~10, (2017) 103508},
  \href{http://arxiv.org/abs/1703.02076}{{\ttfamily arXiv:1703.02076
  [hep-th]}}.

\bibitem{Hartle:2010dq}
J.~Hartle, S.~W. Hawking, and T.~Hertog, ``{Local Observation in Eternal
  inflation},'' \href{http://dx.doi.org/10.1103/PhysRevLett.106.141302}{{\em
  Phys. Rev. Lett.} {\bfseries 106} (2011) 141302},
  \href{http://arxiv.org/abs/1009.2525}{{\ttfamily arXiv:1009.2525 [hep-th]}}.

\bibitem{Hartle:2016tpo}
J.~Hartle and T.~Hertog, ``{One Bubble to Rule Them All},''
  \href{http://dx.doi.org/10.1103/PhysRevD.95.123502}{{\em Phys. Rev. D}
  {\bfseries 95} no.~12, (2017) 123502},
  \href{http://arxiv.org/abs/1604.03580}{{\ttfamily arXiv:1604.03580
  [hep-th]}}.

\bibitem{Almheiri:2019tt}
A.~Almheiri, R.~Mahajan, J.~Maldacena, and Y.~Zhao, ``The page curve of hawking
  radiation from semiclassical geometry,''
  \href{http://arxiv.org/abs/1908.10996}{{\ttfamily 1908.10996}}.
  \url{https://arxiv.org/pdf/1908.10996.pdf}.

\bibitem{Balasubramanian:2020ue}
V.~Balasubramanian, A.~Kar, and T.~Ugajin, ``Islands in de sitter space,''
  \href{http://arxiv.org/abs/2008.05275}{{\ttfamily 2008.05275}}.
  \url{https://arxiv.org/pdf/2008.05275.pdf}.

\bibitem{Shaghoulian:2022aa}
E.~Shaghoulian and L.~Susskind, ``Entanglement in de sitter space,''
  \href{http://arxiv.org/abs/2201.03603}{{\ttfamily 2201.03603}}.
  \url{https://arxiv.org/pdf/2201.03603.pdf}.

\bibitem{Cooper:2018aa}
S.~Cooper, M.~Rozali, B.~Swingle, M.~V. Raamsdonk, C.~Waddell, and D.~Wakeham,
  ``Black hole microstate cosmology,''
  \href{http://arxiv.org/abs/1810.10601}{{\ttfamily 1810.10601}}.
  \url{https://arxiv.org/pdf/1810.10601.pdf}.

\bibitem{Raamsdonk:2021aa}
M.~V. Raamsdonk, ``Cosmology from confinement?,''
  \href{http://arxiv.org/abs/2102.05057}{{\ttfamily 2102.05057}}.
  \url{https://arxiv.org/pdf/2102.05057.pdf}.

\bibitem{Antonini:2022aa}
S.~Antonini, P.~Simidzija, B.~Swingle, and M.~V. Raamsdonk, ``Cosmology from
  the vacuum,'' \href{http://arxiv.org/abs/2203.11220}{{\ttfamily 2203.11220}}.
  \url{https://arxiv.org/pdf/2203.11220.pdf}.

\bibitem{Freivogel_2006}
B.~Freivogel, V.~E. Hubeny, A.~Maloney, R.~C. Myers, M.~Rangamani, and
  S.~Shenker, ``Inflation in {AdS}/{CFT},''
  \href{https://doi.org/10.1088%2F1126-6708%2F2006%2F03%2F007}{{\em Journal of
  High Energy Physics} {\bfseries 2006} no.~03, (Mar, 2006) 007--007}.

\bibitem{Belin_2017}
A.~Belin, J.~de~Boer, J.~Kruthoff, B.~Michel, E.~Shaghoulian, and M.~Shyani,
  ``Universality of sparse d $>$ 2 conformal field theory at large $n$,''
  \href{https://doi.org/10.1007%2Fjhep03%282017%29067}{{\em Journal of High
  Energy Physics} {\bfseries 2017} no.~3, (Mar, 2017) }.

\end{thebibliography}\endgroup

\end{document}